\documentclass[aps,pra,reprint,showpacs,superscriptaddress]{revtex4-1}
\usepackage[english]{babel}
\usepackage{amsmath,amssymb,bbm,graphicx,color,comment,txfonts}
\usepackage[bookmarks=true,colorlinks,citecolor=blue,urlcolor=blue]{hyperref}
\usepackage{dsfont}

\usepackage{setspace} 
\newcommand{\beq}{\begin{equation}}
\newcommand{\eeq}{\end{equation}}
\newcommand{\beqn}{\begin{eqnarray}}
\newcommand{\eeqn}{\end{eqnarray}}

\usepackage{mathtools}
\usepackage{amsmath}
\usepackage{soul}
\usepackage{amssymb}
\usepackage{mathbbol}
\usepackage{xcolor}
\usepackage{graphicx}

\definecolor{dgreen}{rgb}{0.0,0.5,0.0}

\usepackage{epsfig}
\usepackage{bm}	
\renewcommand{\vec}[1]{\bm{#1}}

\DeclarePairedDelimiter\bra{\langle}{\rvert}
\DeclarePairedDelimiter\ket{\lvert}{\rangle}
\DeclarePairedDelimiterX\braket[2]{\langle}{\rangle}{#1 \delimsize\vert #2}
\newcommand{\mean}[1]{\langle #1 \rangle}

\newcommand{\cre}[2]{{#1}^{\dagger}_{#2}}
\newcommand{\ann}[2]{{#1}^{\phantom{\dagger}}_{#2}}

\usepackage{ulem}
\usepackage{cancel,ifthen} 
\newcommand{\cmnt}[2][NoInPuT]{\ifthenelse{\equal{#1}{NoInPuT}}{}{{\color{red}\sout{#1}}} {\color{blue} #2}}

\begin{document}
\normalem	

\title{Signatures of correlated magnetic phases in the two-spin density matrix}

\author{Sebastian Huber}
\affiliation{Physics Department, Arnold Sommerfeld Center for Theoretical Physics, and Center for NanoScience, Ludwig-Maximilians University Munich, Germany}

\author{Fabian Grusdt}
\affiliation{Department of Physics, Harvard University, Cambridge, MA 02138, USA}

\author{Matthias Punk}
\affiliation{Physics Department, Arnold Sommerfeld Center for Theoretical Physics, and Center for NanoScience, Ludwig-Maximilians University Munich, Germany}

\date{\today}

\begin{abstract}

Experiments with quantum gas microscopes have started to explore the antiferromagnetic phase of the two-dimensional Fermi-Hubbard model and effects of doping with holes away from half filling. In this work we show how direct measurements of the system averaged two-spin density matrix and its full counting statistics can be used to identify different correlated magnetic phases with or without long-range order. We discuss examples of phases which are potentially realized in the Hubbard model close to half filling, including antiferrromagnetically ordered insulators and metals, as well as insulating spin-liquids and metals with topological order. For these candidate states we predict the doping- and temperature dependence of local correlators,
which can be directly measured in current experiments.

\end{abstract}

\maketitle

Ultracold atomic gases in optical lattices provide a versatile platform to study strongly correlated phases of matter in a setting with unprecedented control over Hamiltonian parameters \cite{bloch2012quantum,gross2017quantum}. Moreover, the development of quantum gas microscopes now allows for the direct measurement of real space correlation functions with single site resolution in important model systems like the Fermi-Hubbard model, giving access to viable information that can be used to identify various quantum states of matter. Using state of the art technology the many-body wavefunction can now be imaged on a single-site and single-fermion level \cite{edge2015imaging,haller2015single,parsons2016site,boll2016spin,cheuk2016observation} and even the simultaneous detection of spin and charge (i.e. particle-number) degrees of freedom has been achieved \cite{boll2016spin}. In combination with the capability to perform local manipulations, new insights can be obtained into the microscopic properties of strongly correlated quantum many-body systems, which are difficult to access in traditional solid state systems. For example, the hidden string order underlying spin-charge separation in the one-dimensional $t-J$ model has been directly revealed in a quantum gas microscope \cite{hilker2017revealing}. Ultracold atom experiments have also revealed charge ordering in the attractive Fermi-Hubbard model at half filling \cite{mitra2018quantum} and observed longer-ranged antiferromagnetic (AFM) correlations \cite{brown2016observation,mazurenko2017cold}. Furthermore transport properties of the two-dimensional Fermi-Hubbard model were investigated independently for spin and charge degrees of freedom by exposing the system to an external field in the linear response regime \cite{nichols2018spin, brown2018bad}, where clear signatures of bad metal behaviour have been detected in the temperature dependence of the charge conductivity \cite{brown2018bad}. In all these settings, the ultracold atom toolbox can now be applied to gain new insights.

One of the big open problems in the field of strongly correlated electrons is to understand the fate of the AFM Mott insulator in quasi-two-dimensional square lattice systems upon doping it with holes. This problem is particularly relevant in the context of the so-called pseudogap phase in underdoped high-temperature cuprate superconductors \cite{lee2006doping}. In the last decades many works have shown that the two-dimensional one band Hubbard model below half filling captures various phenomena which are found in the phase diagram of cuprates, including superconductivity and charge density wave ordering, among others \cite{lee1992gauge,jarrell2001phase,gull2013superconductivity}.

Quantum gas microscopy experiments are now starting to probe the interesting temperature and doping regime in the Fermi-Hubbard model where correlation effects in doped Mott insulators become visible across the entire system \cite{mazurenko2017cold}, providing valuable insight into this problem. This immediately raises the question how the various symmetric or symmetry broken phases that have been proposed theoretically below half filling can be identified in these experiments. Since accessible temperatures are still rather high, $T \gtrsim 0.5 J$ where $J$ is the super-exchange energy, the correlation length of symmetry broken phases is typically on the order of several lattice spacings, making a direct detection of order parameters challenging. Moreover, various symmetric phases which have been proposed as potential ground-states away from half filling, such as doped resonating-valence bond (RVB) states \cite{anderson1973resonating,anderson1987science,anderson1987resonating}, have very similar short-range spin-spin correlations as magnetically ordered states with a short correlation length. For this reason measurements of spin-spin correlators, which are routinely performed in current quantum gas microscopy experiments, can hardly distinguish these conceptually very different states. In some important cases the symmetric states are characterized by more complicated topological order parameters, which are hard to measure in experiments, however.

\begin{figure}[b!]
\centering
\epsfig{file=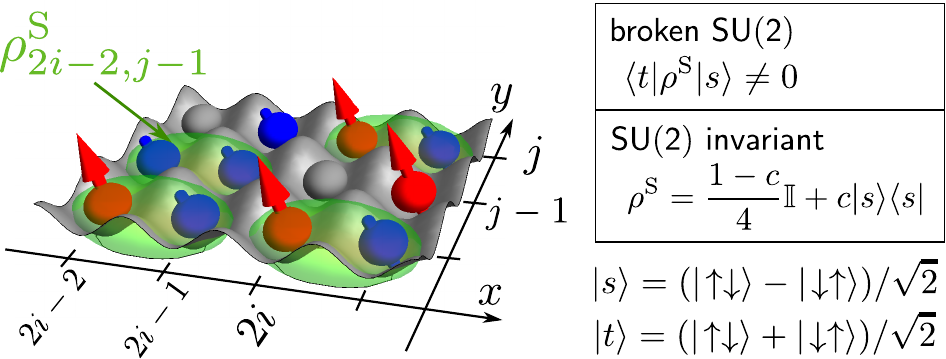, width=0.5\textwidth}
\caption{The two-spin reduced density matrix $\rho^{\rm S}$, measurable in ultracold atom experiments, and its full counting statistics (FCS) can be used to distinguish between symmetric and symmetry broken phases in the Fermi-Hubbard model. The ground state on the square lattice at half filling has AFM order, which leads to non-vanishing singlet-triplet matrix elements $\bra{t} \rho^{\rm S} \ket{s} \neq 0$ as well as a broad distribution of the triplet matrix element in the FCS, even if the correlation length is finite. Below half filling the precise nature of the ground state is still under debate, with doped quantum spin liquids as one possible scenario. These give rise to a ${\rm SU}(2)$ invariant two-spin reduced density matrix with vanishing singlet-triplet matrix elements, as well as a sharp delta-function distribution of the triplet amplitude.} 
\label{figSetup}
\end{figure}

In this work we show how measurements of the reduced two-particle density matrix, see Fig.~\ref{figSetup}, provide a signature of different interesting phases that might be realized in the doped Fermi-Hubbard model at strong coupling. We focus our discussion on phases with strong spin-singlet correlations and show that the presence or absence of SU(2) spin rotation symmetry has a clear signature in the full counting statistics (FCS) of the system-averaged reduced density matrix, allowing to distinguish phases with AFM order from symmetric RVB-like phases, even if the correlation length is finite. In addition we provide results for the doping and temperature dependence of nearest neighbor spin correlators for a metallic antiferromagnet and a doped spin-liquid, as a guide for future experiments.

The paper is organized as follows. In Sec.~\ref{SecFingerprintsQSL} we introduce the two-spin reduced density matrix and discuss how its elements can be measured in quantum gas microscopy experiments. Furthermore, we show how the FCS of the system-averaged reduced density matrix for two neighboring sites can be utilized to distinguish symmetric from symmetry broken phases. The following sections provide explicit examples: in Sec.~\ref{SecSigFHM} we discuss the half filled case and present results for Mott insulators with long-range AFM order as well as for insulating quantum spin liquids. Finally, in Sec.~\ref{SecSigBHF} we calculate the reduced two-spin density matrix and its FCS for two examples below half filling: an AFM metal as well as a metallic state with topological order and no broken symmetries.

\section{ Two-spin reduced density matrix and full counting statistics}
\label{SecFingerprintsQSL}

In this paper we consider the two-spin reduced density matrix of nearest neighbor sites, see Fig,~\ref{figSetup}, which contains information about all local spin correlation functions. We discuss how its matrix elements can be measured in ultracold atom setups and show how states with broken symmetries and long-range order can be distinguished from symmetric states by considering the FCS of the reduced density matrix from repeated experimental realizations. Our approach thus provides tools to address the long-standing question how AFM order is destroyed at finite hole doping using ultracold atom experiments at currently accessible temperatures. 

\subsubsection{Two-spin reduced density matrix}\label{Sec2SpRedDenMat}

The local two-site reduced density matrix $\rho_{\vec{i}, \vec{j}}$, corresponding to sites $\vec{i}$ and $\vec{j}=\vec{i}+\vec{e}_x$ on a square lattice, is defined by tracing out all remaining lattice sites $\vec{r}$ in the environment, $\rho_{\vec{i}, \vec{j}} = {\rm tr}_{\vec{r} \neq \vec{i}, \vec{j}} ~ \rho$, where $\rho$ is the density matrix of the entire system. In general $\rho = e^{- \beta \mathcal{H}}$ describes a thermal state. We consider states with a definite particle number $[ \rho, N ]=0$, where $N$ is the total particle number operator. As a result the two-site density matrix is block diagonal, $\rho_{\vec{i}, \vec{j}} = {\rm diag} (\rho_{\vec{i}, \vec{j}}^{(0)}, \rho_{\vec{i}, \vec{j}}^{(1)},...)$ and contains sectors with $n=0,1,\dots,4$ fermions for spin-1/2 systems (see appendix \ref{AppendixSymRed} for details). In the rest of the paper we will only consider situations where the two sites $\vec{i}$ and $\vec{j}$ are occupied by precisely one fermion each, irrespective of the total fermion density, and calculate the two-spin reduced density matrix $\rho^{\rm S}$. It is obtained from the block with two fermions and proper normalization. Experimentally it can be obtained by post-selecting measurement outcomes with two particles on the two sites.

More specifically we will consider spin-$1/2$ fermions and represent the two-spin reduced density matrix $\rho_{\vec{i}, \vec{j}}^{\rm S}$ in the $z$-basis $\{ \ket{\uparrow \uparrow}, \ket{\uparrow \downarrow}, \ket{\downarrow \uparrow}, \ket{\downarrow \downarrow} \} $, where the first spin refers to site $\vec{i}$ and the second to site $\vec{j}$. It can be written explicitly in terms of local correlation functions in the $z$-basis, 
\begin{align}
\rho^{S}_{\vec{i},\vec{j}}&=\tfrac{1}{4} \mathbbm{1}\label{rhoS1}\\
&+\tfrac{1}{2}
\begin{pmatrix}
\mean{S^{z}_{\vec{i}}}+\mean{S^{z}_{\vec{j}}} & \mean{S^{+}_{\vec{j}}} & \mean{S^{+}_{\vec{i}}} & 0 \\
\mean{S^{-}_{\vec{j}}} & \mean{S^{z}_{\vec{i}}}-\mean{S^{z}_{\vec{j}}} & 0 & \mean{S^{+}_{\vec{i}}} \\
\mean{S^{-}_{\vec{i}}} & 0 & -\mean{S^{z}_{\vec{i}}}+\mean{S^{z}_{\vec{j}}} & \mean{S^{+}_{\vec{j}}} \\
0 & \mean{S^{-}_{\vec{i}}} & \mean{S^{-}_{\vec{j}}} & -\mean{S^{z}_{\vec{i}}}-\mean{S^{z}_{\vec{j}}}
\end{pmatrix}\nonumber\\
&+ \phantom{\tfrac{1}{2}}\begin{pmatrix}
\mean{S^{z}_{\vec{i}} S^{z}_{\vec{j}}} & \phantom{-}\mean{S^{z}_{\vec{i}} S^{+}_{\vec{j}}} & \phantom{-}\mean{S^{+}_{\vec{i}} S^{z}_{\vec{j}}} & \phantom{-}\mean{S^{+}_{\vec{i}} S^{+}_{\vec{j}}} \\
\mean{S^{z}_{\vec{i}} S^{-}_{\vec{j}}} & -\mean{S^{z}_{\vec{i}} S^{z}_{\vec{j}}} & \phantom{-}\mean{S^{+}_{\vec{i}} S^{-}_{\vec{j}}} & -\mean{S^{+}_{\vec{i}} S^{z}_{\vec{j}}} \\
\mean{S^{-}_{\vec{i}} S^{z}_{\vec{j}}} & \phantom{-}\mean{S^{-}_{\vec{i}} S^{+}_{\vec{j}}} & -\mean{S^{z}_{\vec{i}} S^{z}_{\vec{j}}} & -\mean{S^{z}_{\vec{i}} S^{+}_{\vec{j}}} \\
\mean{S^{-}_{\vec{i}} S^{-}_{\vec{j}}} & -\mean{S^{-}_{\vec{i}} S^{z}_{\vec{j}}} & -\mean{S^{z}_{\vec{i}} S^{-}_{\vec{j}}} & \phantom{-}\mean{S^{z}_{\vec{i}} S^{z}_{\vec{j}}}
\end{pmatrix}.\nonumber
\end{align}
Here, $S^\alpha_{\vec{i}}$ is the spin operator on lattice site $\vec{i}$ with $\alpha \in \{ 0,+,-,z \}$ and we define $S^{0}_{\vec{i}} = \mathbb{1}_{\vec{i}}$ as the identity operator. Note that the expectation values $\langle \cdot \rangle$ are defined after post-selecting states with precisely one fermion each on sites $\vec{i}$ and $\vec{j}$. 

For quantum states $\rho$ commuting with $S^z$, i.e. $[\rho,S^z] = 0$, the two-spin density matrix becomes block diagonal. The first two blocks are one-dimensional and correspond to the ferromagnetic basis states $| \! \! \uparrow \uparrow \rangle$ and $| \!\! \downarrow \downarrow \rangle$. The third block corresponds to the two-dimensional subspace spanned by the anti-ferromagnetic states $\ket{\uparrow \downarrow}$ and $\ket{\downarrow \uparrow}$. If the state $\rho$ has an additional $S^z \to -S^z$ symmetry, which follows from a global ${\rm SU}(2)$ symmetry for example, the reduced density matrix simplifies further because the entire second line of Eq.~\eqref{rhoS1} vanishes identically and we get 
\begin{align}
\rho^{S}_{\vec{i},\vec{j}}&=\tfrac{1}{4} \mathbbm{1}\label{rhoS2}\\
&+ \phantom{\tfrac{1}{2}}\begin{pmatrix}
\mean{S^{z}_{\vec{i}} S^{z}_{\vec{j}}} & 0  & 0 & 0 \\
0  & -\mean{S^{z}_{\vec{i}} S^{z}_{\vec{j}}} & \phantom{-}\mean{S^{+}_{\vec{i}} S^{-}_{\vec{j}}} & 0 \\
0 & \phantom{-}\mean{S^{-}_{\vec{i}} S^{+}_{\vec{j}}} & -\mean{S^{z}_{\vec{i}} S^{z}_{\vec{j}}} & 0 \\
0 & 0 & 0 & \phantom{-}\mean{S^{z}_{\vec{i}} S^{z}_{\vec{j}}}
\end{pmatrix}.\nonumber
\end{align}
In this paper we are particularly interested in cases with spontaneously broken or unbroken ${\rm SU}(2)$ symmetry and how it manifests in the two-spin density matrix. For this purpose it is more convenient to represent the two-dimensional sub-block of the reduced density matrix in the singlet-triplet basis defined by
\begin{align}
\ket{s} &= \tfrac{1}{\sqrt{2}}[\ket{\uparrow \downarrow}-\ket{\downarrow \uparrow}],\label{s}\\
\ket{t} &= \tfrac{1}{\sqrt{2}}[\ket{\uparrow \downarrow}+\ket{\downarrow \uparrow}].
\end{align}
In the rest of this paper we will focus on the following combinations of matrix elements of the two-spin density matrix: 
\begin{equation}
p_{\rm f} = \bra{\uparrow \uparrow}~  \rho^{\rm S} ~ \ket{\uparrow \uparrow}  + \bra{\downarrow \downarrow}~  \rho^{\rm S} ~ \ket{\downarrow \downarrow}\label{eqDefpf}
\end{equation}
denotes the probability to observe ferromagnetic correlations on the two sites of interest $\vec{i}$ and $\vec{j}$. It can be directly measured in the $S^z$ basis. Moreover
\begin{equation}
p_{\rm s} = \bra{s} ~ \rho^{\rm S} ~\ket{s} \qquad \text{and} \qquad p_{\rm t}  = \bra{t} ~ \rho^{\rm S} ~\ket{t}\label{eqDefpsAndpt}
\end{equation}
denote the singlet and triplet probabilities and
\begin{multline}
p_{\rm st} = \bra{s} ~ \rho^{\rm S} ~ \ket{t} \\
= \frac{1}{2} \left( \bra{\uparrow \downarrow}~  \rho^{\rm S} ~ \ket{\uparrow \downarrow} - \bra{\downarrow \uparrow}~  \rho^{\rm S} ~ \ket{ \downarrow \uparrow} \right) + i ~  {\rm Im} \bra{\uparrow \downarrow} ~ \rho^{\rm S} ~ \ket{\downarrow \uparrow}
\label{eqDefpst}
\end{multline}
is the singlet-triplet matrix element. The real part of $p_{\rm st}$ can be again directly measured in the $S^z$-basis.

The singlet and triplet probabilities, $p_{\rm s,t}$, can be measured in ultracold atom systems by utilizing the single-site control over spin-exchange interactions in optical superlattices pioneered in Ref. \cite{trotzky2008time}. To this end one can first increase the lattice depth, which switches off all super-exchange interactions. Next a magnetic field gradient along $x$-direction is switched on for a time $\tau_1$ which leads to a Zeeman energy difference $\Delta$ of the two states $\ket{\uparrow \downarrow}$ and $\ket{\downarrow \uparrow}$ and drives singlet-triplet oscillations. Choosing $\tau_1 = \pi / (2 \delta)$ the singlet-triplet basis $\{ \ket{s}, \ket{t} \}$ is mapped to $\{ ( \ket{\uparrow \downarrow} \pm i ~ \ket{\downarrow \uparrow} ) / \sqrt{2} \}$. Subsequently a superlattice can be used to switch on spin-exchange couplings of strength $J$ between sites $(2i,j)$ and $(2i+1,j)$ for a finite time $\tau_2$. By choosing $\tau_2 = \pi / (2 J)$ the original singlet-triplet basis $\{ \ket{s}, \ket{t} \}$ is now mapped on $\{ \ket{\downarrow \uparrow}, \ket{\uparrow \downarrow} \}$. After this mapping a measurement in the $z$-basis directly reveals the singlet and triplet probabilities, $p_{\rm s} = \langle  \ket{\downarrow \uparrow}\bra{\downarrow \uparrow} \rangle$ and $p_{\rm t} = \langle  \ket{\uparrow \downarrow}\bra{\uparrow \downarrow} \rangle$, where the expectation values $\langle \cdot \rangle$ are taken in the measurement basis. 

\subsubsection{Shot-to-shot full counting statistics}

Ultracold atoms not only provide direct access to local correlation functions, but also to the FCS of physical observables, which contain additional information about the underlying many-body states beyond the expectation values in Eq.~\eqref{rhoS1} \cite{hofferberth2008probing}. On the one hand the FCS contain information about quantum fluctuations. On the other hand they can be used to reveal broken symmetries which manifest in long-range order in the system \cite{mazurenko2017cold}. 

In this paper we study the local, reduced two-spin density matrix $\rho^S$ and its FCS in an infinite system. Our goal is to distinguish between fully ${\rm SU}(2)$ symmetric quantum states with short-range correlations, and symmetry broken states with conventional long-range order, despite the fact that these phases can have very similar properties locally. This can be achieved by considering the FCS of $\rho^S$ as follows:
for symmetry broken states the direction of the order parameter varies randomly between experimental shots, giving rise to a specific probability distribution of $\rho^S$ in a given measurement basis. This distribution can be obtained directly from experiments by compiling histograms of a large number of experimental shots. By contrast, this distribution will consist of a single delta-function peak for states with no broken symmetry. It is important to realize, however, that $\rho^S$ also takes different values on different lattice sites within a single experimental shot, which reflects the inherent quantum mechanical probability distribution of $\rho^S$. Determining this quantum mechanical probability distribution is usually referred to as FCS in the condensed matter literature. In order to single out the effect of order parameter fluctuations, we first have to average the two-spin density matrix over the entire system in every shot: 
\begin{equation}
\rho^{\rm S} = \frac{2}{L_x L_y} \sum_{\vec{i} \in {\rm UC}} \rho_{\vec{i}, \vec{i}+\vec{e}_x}^{\rm S},
\label{eqRhoS}
\end{equation}
where $L_{x,y} \to \infty$ denotes the linear system size. We divide the lattice into two-site unit cells along $x$, labeled by one of their site indices $\vec{i} \in {\rm UC}$, in which the reduced two-spin density matrix $\rho_{\vec{i}, \vec{i}+\vec{e}_x}^{\rm S}$ is measured, see Fig.~\ref{figSetup}.  Accordingly, the sum $\sum_{\vec{i} \in {\rm UC}}$ in Eq.~\eqref{eqRhoS} is taken over all such unit-cells. This corresponds to an average over the quantum mechanical probability distribution and ensures that the resulting $\rho^{\rm S}$ is insensitive to quantum fluctuations. Consequently, we can single out effects of the classical probability distribution of $\rho^S$ which arises from different realizations of the order parameter and allows us to distinguish symmetric from symmetry broken states.

The shot-to-shot FCS of $\rho^{\rm S}(n)$ is obtained by measuring $\rho^{\rm S}_{\vec{i}, \vec{i}+\vec{e}_x}$ for all unit cells at positions $\vec{i}$ in a single shot $n$, which yields a measurement outcome for a specified matrix element of $\rho^{\rm S}_{\vec{i}, \vec{i}+\vec{e}_x}(n)$, and taking the system average in Eq.~\eqref{eqRhoS}. This procedure is repeated $N_{\rm s}$ times using a fixed measurement basis (e.g.~$S^z$) and histograms of the matrix elements of $\{ \rho^{\rm S}(n) \}_{n=1...N_{\rm s}}$ yield the desired statistics. 

In a translationally invariant system with short-range correlations the state $\rho$ is symmetric and has no long-range order. In this case the shot-to-shot FCS of $\rho^{\rm S}$ becomes a delta function, 
\begin{equation}
\mathcal{P}[\rho^{\rm S}] |_{\rm sym.} = \delta(\rho^{\rm S} - \rho^{\rm S}_0).
\label{eqPrhoSymm}
\end{equation}
Because of the exponentially decaying correlations, taking the average over the infinite system is equivalent to shot-to-shot averaging of a single pair of spins, $\rho^{\rm S}_0 =  \rho^{\rm S}_{\vec{i}, \vec{i}+\vec{e}_x}(n)$. In a finite system, quantum fluctuations give rise to a distribution peaked around $\rho^{\rm S}_0$ which is expected to have a finite width $w \propto \xi / \sqrt{L_x L_y}$, where $\xi \ll L_{x,y}$ is the finite correlation length.

In a system with a broken symmetry and long-range correlations extending over the entire system, in contrast, spatial and shot-to-shot averaging are not equivalent in general. All measurement outcomes $\rho^{\rm S}_{\vec{i}, \vec{i}+\vec{e}_x}(n) \equiv \rho^{\rm S}_{\vec{i}, \vec{i}+\vec{e}_x}(n,\vec{\Omega}(n))$ explicitly depend on the order parameter $\Omega(n)$ associated with the long-range correlations in the system for shot $n$. As a result the system-averaged two-spin density matrix $\rho^{\rm S}(n) \equiv \rho^{\rm S}(n,\vec{\Omega}(n))$ explicitly depends on the order parameter $\vec{\Omega}(n)$. 

In systems with spontaneous symmetry breaking, the order parameter $\vec{\Omega}(n)$ fluctuates from shot to shot. Because the averaging over the infinite system in Eq.~\eqref{eqRhoS} makes $\rho^{\rm S}(n)$ insensitive to local quantum fluctuations, it only depends on the order parameter, i.e. $\rho^{\rm S}(n) \equiv \rho^{\rm S}_0(\vec{\Omega}(n))$. Therefore the shot-to-shot FCS of $\rho^{\rm S}$ reflects the probability distribution $\mathcal{P}[\vec{\Omega}]$ of the order parameter $\vec{\Omega}(n)$. The probability distribution of the system-averaged reduced density matrix thus takes the form
\begin{equation}
\mathcal{P}[\rho^{\rm S}] |_{\rm sym. broken} = \int d\vec{\Omega}  ~ \mathcal{P}[\vec{\Omega}] ~ \delta(\rho^{\rm S} - \rho^{\rm S}_0(\vec{\Omega})).
\label{eqPrhoSymmBroken}
\end{equation}
When the order parameter $\vec{\Omega}(n)$ takes a different value in every shot $n$, the reduced two-spin density matrix is characterized by a broad distribution function in general. Its width $w$ converges to a finite value in the limit of infinite system size. 

The reduced density matrix $\rho^S$ defined on neighboring sites $\vec{i}$ and $\vec{i}+\vec{e}_x$, forming two-site unit-cells of the square lattice, is sensitive to order parameters indicating spontaneously broken ${\rm SU(2)}$ symmetries, either ferromagnetic or AFM, and some discrete translational symmetries as expected for valence bond solids (VBS). We note, however, that $\rho^S$ is insensitive to other order parameters. In such cases the distribution function becomes narrow, as in Eq.~\eqref{eqPrhoSymm}, and the underlying ordering cannot be detected.

We close this section by a discussion of finite temperature effects in the two-dimensional Fermi-Hubbard model. Due to the Mermin-Wagner theorem \cite{mermin1966absence}, no true long-range order can exist at non-zero temperatures, and the ${\rm SU(2)}$ symmetry remains unbroken. However, the correlation length increases exponentially with decreasing temperatures \cite{mazurenko2017cold}, until it reaches the finite system size. In this case, the state cannot be distinguished from a symmetry-broken state, and from Eq.~\eqref{eqPrhoSymmBroken} we expect broad distribution functions of the entries in the two-spin reduced density matrix. Because of the finite system size, the averaging in Eq.~\eqref{eqRhoS} does not eliminate all quantum fluctuations, however, which leads to broadened distribution functions; see Refs. \cite{mazurenko2017cold,Humeniuk2017} for explicit calculations. When the system is too small, a clear distinction between $\rm SU(2)$-broken and ${\rm SU}(2)$-symmetric phases is no longer possible.

\section{Two-spin density matrix and full counting statistics at half filling}
\label{SecSigFHM}

As described above, the shot-to-shot FCS of the reduced two-spin density matrix can be used to distinguish states with broken symmetries from symmetric states. Here we consider two important examples at half filling: an AFM Mott insulator and an insulating spin liquid in the two-dimensional square lattice Fermi-Hubbard model. We emphasize that the ground state is known to be an AFM in this case. The main purpose of the spin liquid example is to highlight the stark contrast between a magnetically ordered and a symmetric state in the two-particle density matrix in order to set the stage for the discussion of systems below half filling. The FCS of system averaged local observables is a very sensitive probe to distinguish ordered from disordered states, which are particularly hard to discern if the correlation length is short.

The ground state of the Fermi-Hubbard model at half filling breaks the ${\rm SU}(2)$ spin-rotation symmetry, it has long-range AFM order and it is invariant under translations by integer multiples of $\vec{e}_x \pm \vec{e}_y$. The corresponding order parameter is given by the staggered magnetization, $\vec{\Omega} = (-1)^{j_x+j_y} ~ \langle \vec{S}_{j_x,j_y} \rangle$. Because $\vec{\Omega}$ points in a different direction in every experimental realization and the spins are always measured in the $S^z$ basis, we expect a broad distribution of the reduced two-spin density matrix between different experimental shots. 

For example, the real part of the system-averaged singlet-triplet matrix element is given by the staggered magnetization,
\begin{eqnarray}
{\rm Re} p_{\rm s,t} &=&  \frac{1}{L_x L_y} \sum_{\vec{i} \in {\rm UC}} \left(  \bra{\uparrow \downarrow}~  \rho^{\rm S}_{\vec{i},\vec{i}+\vec{e}_x} ~ \ket{\uparrow \downarrow} - \bra{\downarrow \uparrow}~  \rho^{\rm S}_{\vec{i},\vec{i}+\vec{e}_x} ~ \ket{ \downarrow \uparrow}   \right) \notag \\
&=&  \frac{1}{L_x L_y} \sum_{\vec{r}} (-1)^{r_x,r_y} ~ \langle S^z_{\vec{r}} \rangle \equiv M^z_{\rm stag},
\end{eqnarray}
see Eq.~\eqref{eqDefpst}. Note that the sum $\sum_{\vec{i} \in {\rm UC}}$ in the first line is taken over all two-site unit-cells, whereas the sum $\sum_{\vec{r}}$ in the second line extends over all lattice sites. The distribution function $\mathcal{P}[M^z_{\rm stag}]$ of the staggered magnetization has been measured in a finite-size system using ultracold fermions \cite{mazurenko2017cold}. At low temperatures $\mathcal{P}[M^z_{\rm stag}]$ becomes a broad distribution which approaches a box-like shape for an infinite system at zero temperature $T=0$ \cite{Humeniuk2017,KanaszNagy2018}. In contrast, a narrow distribution would be expected for a ${\rm SU}(2)$ invariant quantum spin liquid. 

\begin{figure}[t!]
\centering
\epsfig{file=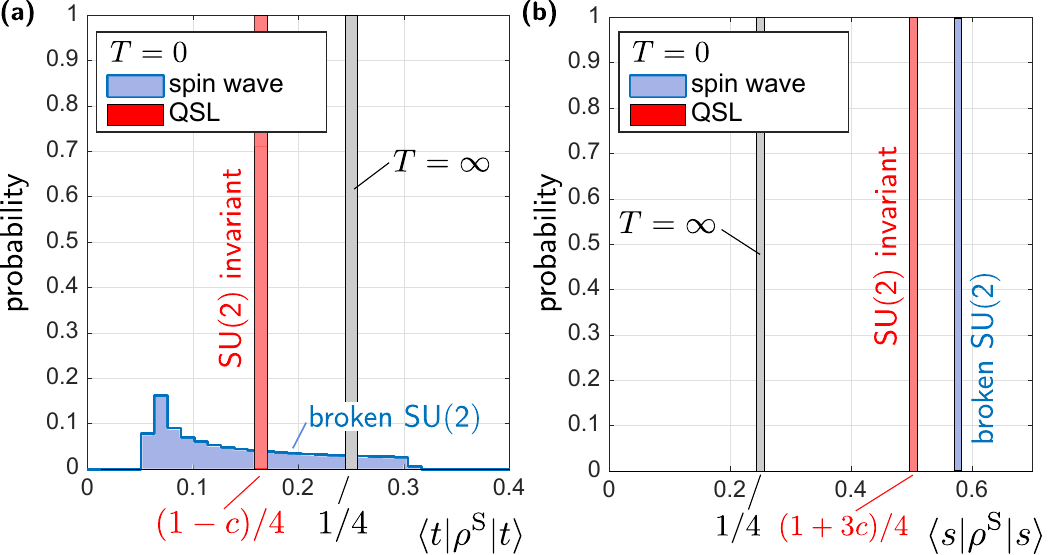, width=0.5\textwidth}
\caption{Fingerprints for the spontaneous breaking of ${\rm SU}(2)$ invariance in the shot-to-shot FCS of the system-averaged reduced two-spin density matrix $\rho^{\rm S}$, see Eq.~\eqref{eqRhoS}. When the ${\rm SU}(2)$ symmetry is spontaneously broken, the order parameter points in a different direction in every shot. This results in a broad distribution function of some of the matrix elements of $\rho^{S}$. (a) We use spin-wave theory to calculate the distribution of the triplet matrix element $\bra{t} \rho^{\rm S} \ket{t}$ in (a) for an infinite Heisenberg AFM at half filling. For a ${\rm SU}(2)$ invariant quantum spin liquid the distribution function in an infinite system becomes a delta peak. (b) The distribution of the singlet matrix element $\bra{s} \rho^{\rm S} \ket{s}$ is a delta peak in an infinite system even when the ${\rm SU}(2)$ symmetry is broken, because the singlet state $\ket{s}$ itself is ${\rm SU}(2)$ invariant.} 
\label{figCoarseGrainedFCS}
\end{figure}

In Fig.~\ref{figCoarseGrainedFCS} (a) we present the shot-to-shot FCS of the triplet probability $p_{\rm t}$. For an infinite system at $T = \infty$, i.e. $\rho^{\rm S} = \mathbb{1}/4$, and $\mathcal{P}[p_{\rm t}] = \delta(p_{\rm t} - 1/4)$ becomes a delta-function at $p_{\rm t}=1/4$. For the quantum Heisenberg AFM at zero temperature, $T=0$, we predict a broad distribution with $0.05 \leq p_{\rm t} \leq 0.3$. Our calculation uses linear spin wave theory to obtain the two-spin density matrix $\rho^{\rm S}(\vec{\Omega} \propto \vec{e}_z)$ in a basis where the AFM order $\vec{\Omega}$ points in $z$-direction, as discussed in detail in the appendix \ref{AppendixHolPrim}. We obtain the shot-to-shot FCS by sampling random directions $\vec{\Omega}$ and performing basis transformations $U(\vec{\Omega})$ rotating the AFM order to point along $\vec{\Omega}$, i.e. $\rho^{\rm S}(\vec{\Omega}) = U^\dagger(\vec{\Omega}) ~ \rho^{\rm S}(\vec{e}_z) ~ U(\vec{\Omega})$.

We extend the spin wave calculation by an external staggered magnetic field with strength $h_z$ as a control parameter for quantum fluctuations. For a strong magnetic field the ground state is close to a classical N\'eel configuration and the FCS of the system-average triplet matrix element is box-like shaped between $0 \leq \bra{t} \rho^S \ket{t} \leq 0.5$ (see appendix \ref{AppendixHolPrim} for details). When we reduce the strength of the external magnetic field $h_z$ we find that the probability distribution $\mathcal{P}[p_t]$ develops an onset at finite $p_t > 0$, as shown for $h_z=0$ in Fig. \ref{figCoarseGrainedFCS} (a). We conclude that this characteristic onset is due to quantum fluctuations, which are suppressed when $h_z$ is large.

For a ${\rm SU}(2)$ invariant quantum spin liquid at $T=0$ the system-averaged triplet probability $p_{\rm t}$ is a delta distribution, i.e. $\mathcal{P}[p_{\rm t}] = \delta(p_{\rm t} - p_{\rm t}^0)$. In contrast to the infinite temperature case, the expectation value $p_{\rm t}^0$ generically takes values different from $1/4$, however. The ${\rm SU}(2)$ invariance determines the form of the two-spin density matrix to be \cite{roy2016response}, 
\begin{align}
\rho^{S}=\frac{1-c}{4} \mathbbm{1} + c \ket{ s }\bra{ s }, 
\label{eqHFRDM1}
\end{align}
up to a non-universal number $c \in [0,1]$, which parametrizes the strength of singlet correlations. The triplet probability is thus given by $p_{\rm t} = (1-c)/4$.

In Fig.~\ref{figCoarseGrainedFCS} (b) we show the shot-to-shot FCS of the singlet probability $p_{\rm s}$, averaged over an infinite system. Because the singlet state $\ket{s}$ is invariant under ${\rm SU}(2)$ transformations, $p_{\rm s}$ is independent of the order parameter $\vec{\Omega}$ even when the ${\rm SU}(2)$ symmetry is spontaneously broken. As a result we obtain delta distributions $\mathcal{P}[p_{\rm s}] = \delta(p_{\rm s} - p_{\rm s}^0)$ in all considered scenarios. For a system at infinite temperature $p_{\rm s}^0=1/4$, i.e. $c=0$ in Eq.~\eqref{eqHFRDM1}, for a quantum spin liquid with the two-spin density matrix in Eq.~\eqref{eqHFRDM1} $p_{\rm s}^0=(1+3 c)/4$, and for the quantum Heisenberg AFM with broken ${\rm SU}(2)$ symmetry we find from a linear spin-wave calculation that $p_{\rm s}^0|_{\rm AFM}=0.57$.

Analyzing the shot-to-shot FCS of the reduced two-spin density matrix represents a powerful method to distinguish states with long-range correlations from symmetric quantum spin liquids. Although we only discussed a spontaneously broken ${\rm SU}(2)$ symmetry at half filling so far, the approach also allows to distinguish symmetry broken states at finite doping from fully symmetric states. Valence-bond solids, which are fully ${\rm SU}(2)$ invariant but spontaneously break the lattice translation symmetry, can also be identified in this way, which is of particular importance for frustrated quantum magnets as in the $J_1-J_2$ model \cite{wang2016tensor,jiang2012spin}.

\section{Two-spin density matrix and full counting statistics below half filling}
\label{SecSigBHF}

One of the big open questions in studies of the Fermi-Hubbard model is to determine the nature of the ground-state for strong interactions slightly below half filling. This is the regime where the infamous metallic pseudogap phase has been observed in cuprate high-temperature superconductors \cite{lee2006doping}, the main properties of which are believed to be captured by the Fermi-Hubbard model \cite{ferrero2009pseudogap,gull2010momentum,gull2013superconductivity}, even though controlled, reliable numerical results do not exist. Quantum gas microscopy experiments have started to probe this regime and might provide valuable insight into this problem \cite{mazurenko2017cold}. While many different scenarios have been proposed theoretically to explain the pseudogap phenomenology in the cuprates, we will focus our discussion of the reduced two-spin density matrix below half filling on two possible phases, in close analogy to the half filled case. The first example is a simple metallic state with long-range AFM order, whereas the second example describes a so-called fractionalized Fermi liquid (FL*), which can be understood as a doped quantum spin liquid with topological order and no broken symmetries \cite{senthil2003fractionalized}. In particular we are going to highlight signatures of these two phases as a function of temperature and as a function of the density of doped holes away from half filling. It is important to emphasise that we always consider the two-spin reduced density matrix for two neighbouring, singly occupied sites. Experimentally, this requires a post-selection of realizations where each of the two lattice sites in question is occupied by a single atom.

In the following we compute the reduced two-spin density matrix for AFM metals and FL* using a slave-particle approach introduced by Ribeiro and Wen \cite{Ribeiro2006}. This approach is quite versatile and allows to describe a variety of different possible phases in the $t-J$ model, which provides an effective description of the Fermi-Hubbard model in the large $U$ limit \cite{Ribeiro2006,Punk2012,mei2012luttinger}. It is important to emphasize, however, that this approach is not quantitatively reliable. Its strength is to provide qualitative predictions for different phases that might be realized in the $t-J$ model. The stability of analogous slave-particle mean-field ground states has been discussed e.g.~in Refs.~\cite{hermele2004stability,wen1991mean}. In the following we briefly summarize the main idea and refer to the appendix \ref{AppendixDopedCarrier} for a detailed discussion. 

Our starting point is the $t-J$ Hamiltonian 
\begin{align}
H_{tJ}=J&\sum_{\mean{ij}\in NN}(\bold{S}_{i} \cdot \bold{S}_{j}-\tfrac{1}{4}\mathcal{P}n_{i,\sigma}n_{j,\sigma}\mathcal{P})\label{tJHam}\\
&+t\sum_{\mean{ij}\in NN}\mathcal{P}(\cre{c}{i,\sigma}\ann{c}{j,\sigma}+\cre{c}{j,\sigma}\ann{c}{i,\sigma})\mathcal{P},\nonumber
\end{align}
where we restrict to nearest neighbour hopping as in ultracold atom experiments. Here, the spin operator $\bold{S}_{i}$ is given in terms of Gutzwiller projected fermion operators as $\bold{S}_{i}=\tfrac{1}{2}\mathcal{P}\cre{c}{i,\alpha}\sigma_{\alpha,\beta}\ann{c}{i,\beta}\mathcal{P}$, where $\mathcal{P}$ is the Gutzwiller projector which projects out doubly occupied sites.

The main idea of the slave particle description of Ribeiro and Wen is to introduce two degrees of freedom per lattice site: one localized spin-1/2 (represented by the operator $\bold{\tilde{S}}_i$), as well as one charged spin-1/2 fermion described by fermionic operators $ \cre{d}{i,\sigma}, \ann{d}{i,\sigma}$, which represents a doped hole (referred to as dopon in the following). The three physical basis states per lattice site of the $t-J$ model are then related to the two slave-degrees of freedom per lattice site via the mapping
\begin{align}
\ket{\uparrow}_i &\leftrightarrow \ket{\uparrow 0}_i,\label{SpinonDoponMap1}\\
\ket{\downarrow}_i &\leftrightarrow \ket{\downarrow 0}_i,\\
\ket{0}_i &\leftrightarrow \frac{\ket{\uparrow \downarrow}_i-\ket{\downarrow \uparrow}_i}{\sqrt{2}},\label{SpinonDoponMap3}
\end{align}
i.e.~a physical hole is represented by a spin-singlet of a localized spin and a dopon. Other states of the enlarged slave-particle Hilbert space, such as the triplet states $\ket{\uparrow \uparrow}$, $\ket{\downarrow \downarrow}$, $\frac{\ket{\uparrow \downarrow}_i+\ket{\downarrow \uparrow}_i}{\sqrt{2}}$ and doubly occupied dopon states are unphysical and need to be projected out. 

In terms of the slave-particle degrees of freedom, the Gutzwiller projected electron operator of the $t-J$ model takes the form
\begin{align}
 \cre{\tilde{c}}{i,\alpha}=\mathcal{P}\cre{c}{i,\alpha}\mathcal{P}=s_{\sigma}\tfrac{1}{\sqrt{2}}\mathcal{\tilde{P}}[(\tfrac{1}{2}+s_{\sigma}\tilde{S}^{z}_{i})- \tilde{S}^{s_{\sigma}}_{i} \ann{d}{i,\sigma}]\mathcal{\tilde{P}},\label{electronOperatorSD}
\end{align}
where $\tilde{\mathcal{P}}$ projects out doubly occupied dopon sites and $s_{\uparrow / \downarrow}=\pm1$. In this slave-partice representation the $t-J$ Hamiltonian in Eq. \eqref{tJHam} takes the form
\begin{align}
H^{d}_{tJ}=H^{d}_{J}+H^{d}_{t},\label{dtJHam}
\end{align}
where
\begin{widetext}
\begin{align}
H^{d}_{J}=&J\sum_{\mean{ij} \in NN}(\bold{\tilde{S}}_{i}\bold{\tilde{S}}_{j}-\tfrac{1}{4})\mathcal{\tilde{P}}(1-\cre{d}{i}\ann{d}{i})(1-\cre{d}{j}\ann{d}{j})\mathcal{\tilde{P}},\label{dJHam}\\
H^{d}_{t}=&\frac{t}{2} \sum_{\mean{ij}\in NN} \mathcal{\tilde{P}} [(\cre{d}{i}\vec{\sigma}\ann{d}{j})\cdot(i \bold{\tilde{S}}_{i} \times \bold{\tilde{S}}_{j}-\frac{\bold{\tilde{S}}_{i}+\bold{\tilde{S}}_{j}}{2})+\tfrac{1}{4}\cre{d}{i}\ann{d}{j}+\cre{d}{i}\ann{d}{j}\bold{\tilde{S}}_{i}\bold{\tilde{S}}_{j}+h.c.] \mathcal{\tilde{P}}.\label{dtHam}
\end{align}
\end{widetext}
One big advantage of this approach is that the Hamiltonian \eqref{dtJHam} does not mix the physical states with the unphysical triplet states in the enlarged Hilbert space. A projection to the physical states in the enlarged Hilbert space is thus not necessary. Note that the Hamiltonian \eqref{dtJHam} resembles a Kondo-Heisenberg model of localized spins $\tilde{\bold{S}}_i$ interacting with a band of itinerant spin-1/2 fermions $ \cre{d}{i,\sigma}, \ann{d}{i,\sigma}$, which describe the motion of doped holes.
The density $p$ of doped holes away from half filling in the $t-J$ model equals the density of dopons in the slave-particle description, $p=\frac{1}{N}\sum_{i}\mean{\cre{d}{i}\ann{d}{i}}$. We conclude that in the low doping regime, where the density of dopons is very small, the Gutzwiller projector $\tilde{\mathcal{P}}$ for the dopons can be neglected. In the same regime the exchange interaction between spins in Eq. \eqref{dJHam} is just renormalized  by the presence of dopons and can be approximated as $J\mathcal{\tilde{P}}(1-\cre{d}{i}\ann{d}{i})(1-\cre{d}{j}\ann{d}{j})\mathcal{\tilde{P}}\approx J (1-p)^2 \equiv \tilde{J}$. The second part of the Hamiltonan $H^{d}_{t}$ describes the hopping of dopons as well as their interaction with the localised spins.

The two phases of interest in this section can now be understood as follows: in the AFM metal the localized spins $\bold{\tilde{S}}_i$ as well as the doped spins order AFM and the dopons form a Fermi-liquid on the background of ordered spins. By contrast, the FL* corresponds to a phase where the localized spins are in a spin-liquid state, and the dopons form a Fermi-liquid on top \cite{senthil2003fractionalized,Punk2012}. The absence of magnetic order requires frustrated spin-spin interactions, which in this case can arise from RKKY-interactions mediated by the dopons. Note that the FL* state violates the conventional Luttinger theorem \cite{oshikawa2000topological}, which states that the volume enclosed by the Fermi surface in an ordinary metal without broken symmetries is proportional to the total density of electrons (or holes) in the conduction band. Instead, the FL* state has a small Fermi surface with an enclosed volume determined by the density of doped holes away from half filling ($p$), rather than the full density of holes measured from the filled band $(1+p)$ \cite{senthil2004weak}. It has been argued that such an FL* state shares many properties with the pseudogap phase in underdoped cuprates \cite{kaul2007algebraic,qi2010effective,mei2012luttinger,Punk2012,sachdev2016novel,punk2015quantum,huber2018electron,feldmeier2018exact}.

\begin{figure*}[t!]
\centering\
\epsfig{file=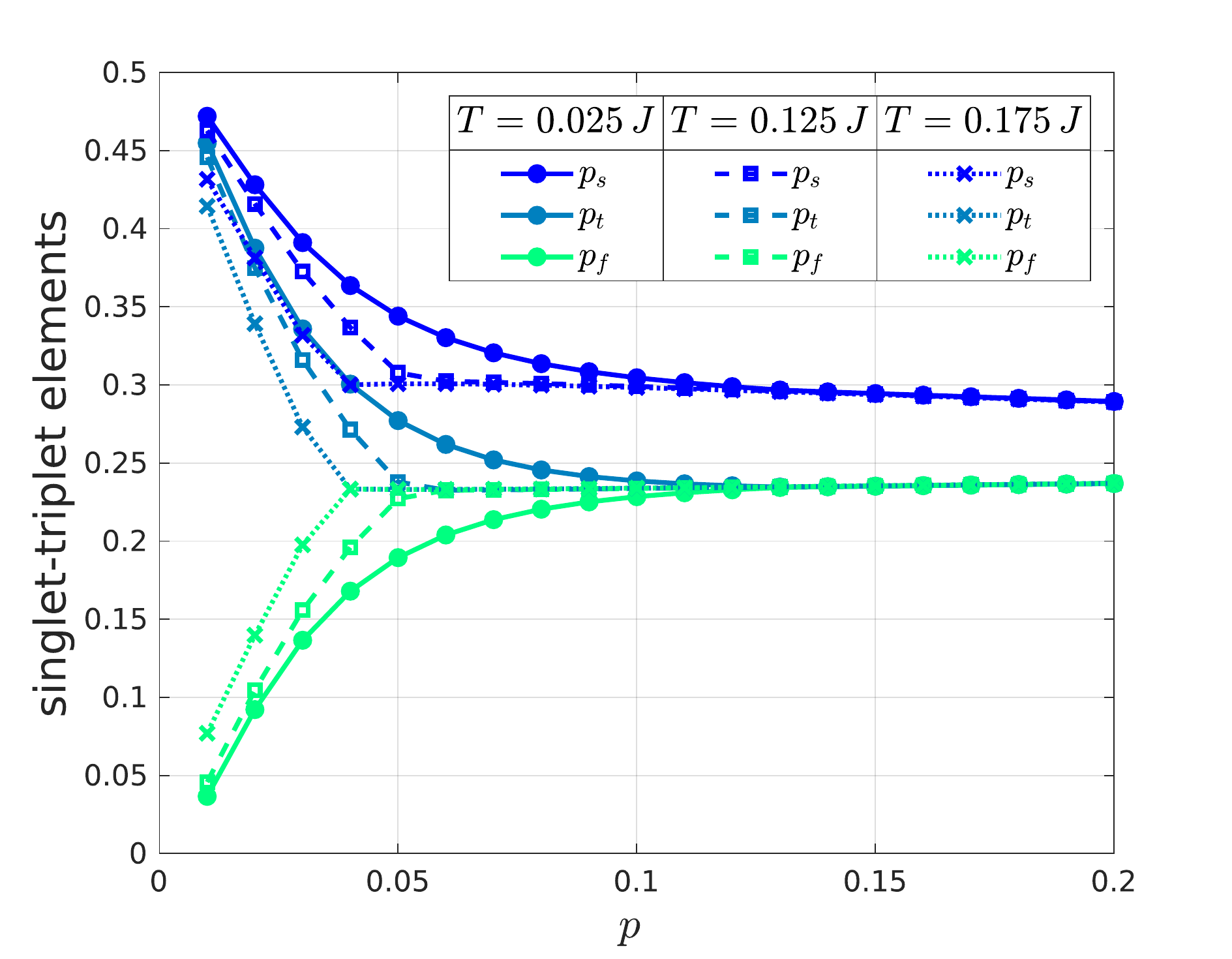, width=0.48\textwidth}
\epsfig{file=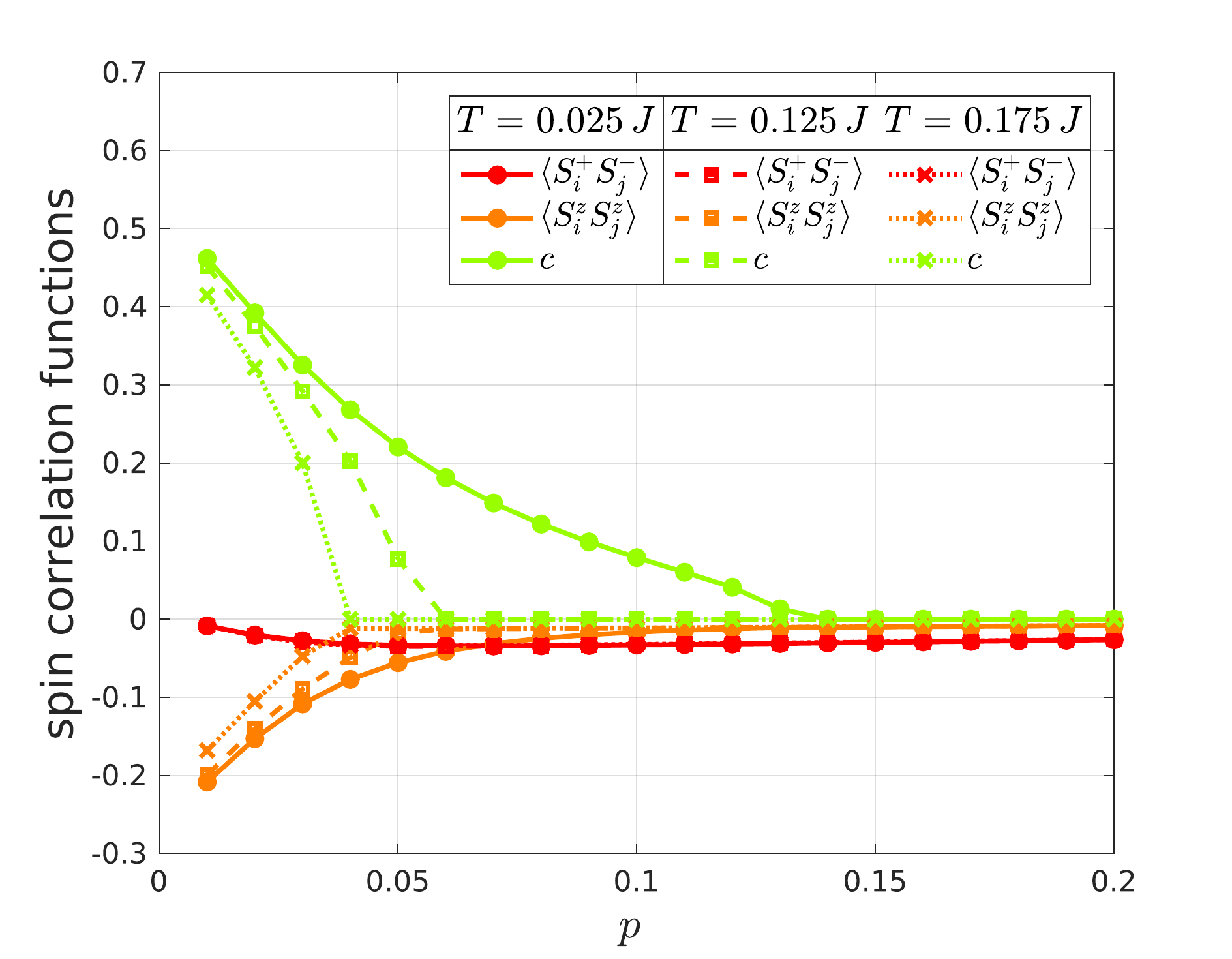, width=0.48\textwidth}
\caption{Doping dependence of local correlations in the AFM metal. Part (a) of this figure shows the short-range singlet, triplet and ferromagnetic amplitudes ($p_s$, $p_t$ and $p_f$) as a function of doping $p$. We show here data for three temperatures, namely $T = 0.025 J$ (bold line), $T = 0.125J$ (bold dashed) and $T=0.175J$ (dotted). The amplitudes decrease with increasing temperature due to larger thermal fluctuations. The triplet and ferromagnetic amplitudes approach each other when doping the system continuously from $p = 0.01$ to $p= 0.2$. The singlet amplitude $p_s$ {remains larger than} the other two amplitudes {at large doping, by an amount determined by $c_{p \gtrsim 0.1} \approx 0.07$} in Eq. \eqref{eqHFRDM1}. The order parameter $\mean{S_i^z}$ in part (b) indicates that the ${\rm SU(2)}$ symmetry is restored at a critical doping of $p \approx 0.06$ for the temperature $T=0.125J$. 
}
\label{figAFMFiniteHoleDoping}
\end{figure*}

In order to obtain a phenomenological description of the above mentioned phases we follow Ribeiro and Wen and employ a slave-fermion description of the localized spins 
\begin{equation}
\tilde{\bold{S}}_i = f^\dagger_{i \alpha} \vec{\sigma}_{\alpha \beta} f_{i \beta},
\end{equation}
where $f_{i\alpha}$ and $f^\dagger_{i\alpha}$ are canonical spin-1/2 fermion operators. This description is particularly suited to construct spin-liquid states of the localized spins, where the $f$ fermions describe spinon excitations, as will be discussed later. In the following we introduce the mean-fields used to decouple the various interaction terms in the Hamiltonian, and which are the basis of a phenomenological description of the two phases mentioned above.

\begin{figure*}[t!]
\centering\
\epsfig{file=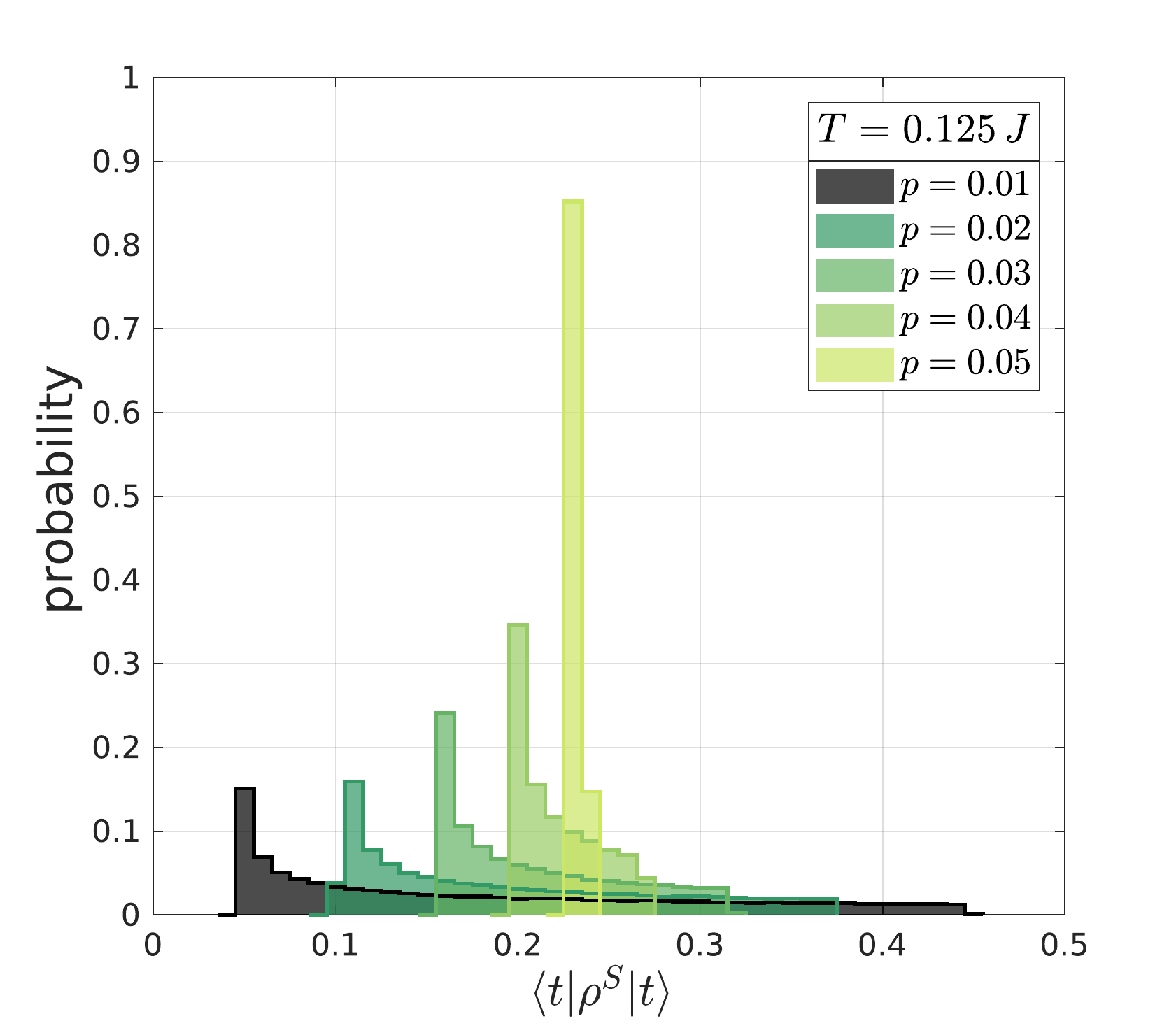, width=0.48\textwidth}
\epsfig{file=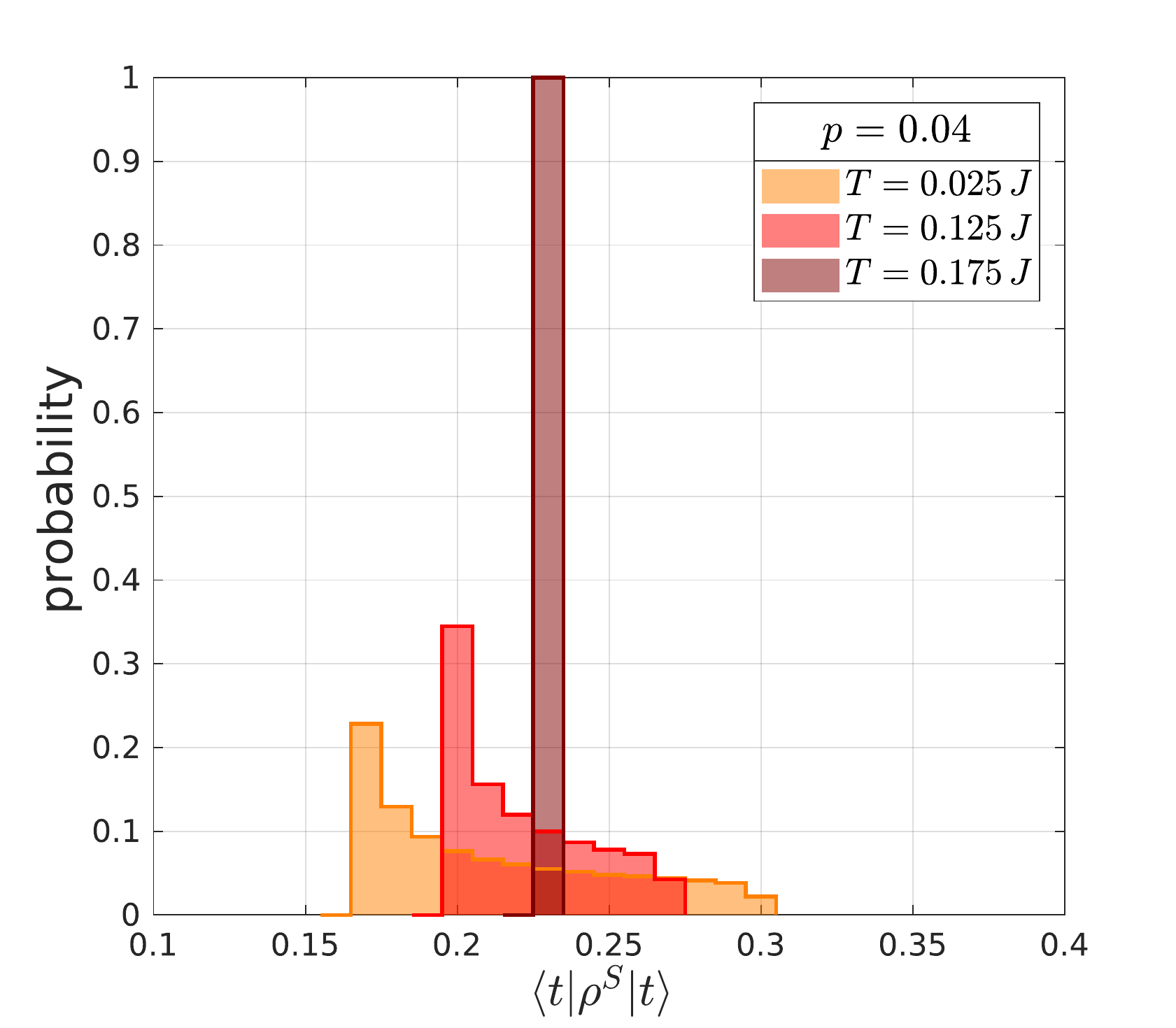, width=0.48\textwidth}
\caption{Full counting statistics of local correlations in the AFM metal. We show the FCS of the system-averaged triplet matrix element $\bra{t} \rho^S \ket{t}$ of the two-spin density matrix. The left panel shows the FCS distribution at a fixed temperature $T=0.125J$ for different doping levels $p$. It narrows as function of $p$ and turns into a sharp peak at the critical doping $p \approx 0.05$, where the SU(2) symmetry is restored. The sharp peak at a value around $0.23$ determines the parameter $c$ in the density matrix in Eq.~\eqref{eqHFRDM1}, which takes the value $c_{p \approx 0.05} \approx 0.07$. The right panel shows a similar behavior of the probability distribution as function of temperature at fixed doping, where thermal fluctuations restore the SU(2) symmetry and lead to a narrow peak around a value $\bra{t} \rho^S \ket{t} = 0.23$.} 
\label{figAFM_FCS}
\end{figure*}

\subsection{Antiferromagnetic metal}
Let us focus on the first part of the Hamiltonian in Eq. \eqref{dJHam} describing a renormalized spin exchange interaction. Here we use a mean-field decoupling which allows for an effective hopping and a pairing amplitude for the spinons:
\begin{align}
\chi_{ij} &= \mean{\cre{f}{i,\alpha}\ann{f}{j,\beta}}\delta_{\alpha\beta}.\label{MF_chi} \\
\ann{\Delta}{ij} &= \mean{\ann{f}{i,\alpha}\ann{f}{j,\beta}}\epsilon_{\alpha\beta},\label{MF_delta}
\end{align}
Note that the spinon pairing amplitude $\Delta_{ij}$ effectively accounts for singlet correlations between the localised spins
and we assume that it has a d-wave form, where $\Delta = \pm\mean{\Delta_{ij}}$ for $j=i+\hat{e}_{x/y}$. 

The second part of the Hamiltonian in Eq. \eqref{dtHam} consists of terms describing the hopping of dopons as well as their interaction with localized spins. We now discuss the effect of each of these terms using a mean-field analysis. Since we assume that spins have local AFM correlations, the cross product between spin operators $\mean{\bold{\tilde{S}}_{i} \times \bold{\tilde{S}}_{j}}$ vanishes for nearest neighbors. The last term in Eq. \eqref{dtHam} can be decoupled either to generate an effective hopping of spinons and/or dopons. In the latter case the dopons are considered to hop in a locally N\'eel ordered background, i.e. $4\mean{\bold{\tilde{S}}_{i}\bold{\tilde{S}}_{j}} \approx (-1)^{i_x -  j_x +  i_y - j_y}$. This however cancels with the third term in Eq.~\eqref{dtHam} and thus effectively leads to a vanishing dispersion for the dopons. In contrast numerical and theoretical studies for a single hole described by the $t-J$ model show a dispersion relation with a minimum around $(\pi/2,\pi/2)$ \cite{kim1998systematics,martinez1991spin,liu1992dynamical,dagotto1994flat,tohyama2000angle}. To overcome this discrepancy we allow for further neighbor hopping amplitudes $\{ t_1, t_2, t_3 \}$, so that dopons can effectively tunnel up to second and third nearest neighbour sites within our mean-field analysis. The nearest neighbour hopping amplitude is thereby set to $t_1 = t$. As motivated in the work by Ribeiro and Wen, the second and third nearest neighbour hopping amplitudes scale approximately as $t_2 = 2 t_3 \approx J$. 

Finally the second term of the Hamiltonian in Eq. \eqref{dtHam} plays a major role here, since it takes the form of a Kondo coupling between the dopons and the spins. The resonances of the according processes are thus significantly larger in case of a strongly developed spin ordered background. In order to describe such a macroscopically developed AFM spin background we introduce the following mean-field amplitudes
\begin{align}
m^z &= (-1)^{i_x + i_y} \frac{1}{2} \mean{\cre{f}{i,\alpha} \ann{f}{i,\beta}} \sigma^z_{\alpha \beta},\label{MF_mz}\\
n^z &= -(-1)^{i_x + i_y} \sum_{\nu \in \{ 2,3\} } \frac{t_{\nu}}{8} \sum_{\hat{u}_v} \mean{\cre{d}{i,\alpha} \ann{d}{i+\hat{u}_v,\beta}} \sigma^z_{\alpha \beta}. \label{MF_nz}
\end{align}
The dopon magnetization $n^z$ measures thereby the net effect of a hole with respect to an AFM ordered background. The first sum runs over further neighbors $v=2,3$, whereas the second sum includes the following contributions $\hat{u}_2 = \pm e_{\hat{x} / \hat{y}} \pm e_{\hat{y} / \hat{x}}$ and $\hat{u}_3 = \pm  2e_{\hat{x} / \hat{y}}$.
The detailed analysis of the mean-field self-consistency equations is part of appendix \ref{AppendixDopedCarrier} and follows closely Ref. \cite{Ribeiro2006}. Note that we do not include a hybridization between spinons and dopons, i.e.~mean fields of the form $\langle  f^\dagger_i d_i \rangle$. Such terms are only important for a description of the ordinary Fermi liquid at large doping. 

In Fig. \ref{figAFMFiniteHoleDoping} we show the self-consistent mean-field results for $t=16 J$. Note that we choose such a relatively small value of $J$ because the mean-field computation overestimates the extent of the AFM phase, which is known to vanish at a few percent doping in realistic situations. A small value of $J$ reduces the extent of the AFM phase as function of doping and thus allows us to compensate for this artifact of mean-field theory. The AFM order parameter as function of doping for three different temperatures is shown in Fig.~\ref{figAFMFiniteHoleDoping}b, together with the nearest neighbor spin correlators.
The short-range singlet and triplet probabilities $p_s$, $p_t$ for pairs of nearest neighbor sites are shown in Fig. \ref{figAFMFiniteHoleDoping} (a). Both are close to the value $p_{s/t} \approx 0.5$ at half filling, and decrease with doping. At higher temperatures, thermal fluctuations reduce the absolute values of the amplitudes. Beyond a temperature dependent threshold between $p \approx 0.05$ and $p \approx 0.1$ the ferromagnetic and triplet amplitude are very close to each other with a value of $p_{t/f} \approx 0.24$, indicating that the SU(2) symmetry is restored. In this regime, the singlet probability $p_s$ deviates from $p_t = p_f$ by an amount which is related to the constant $c$ characterizing the ${\rm SU(2)}$ invariant two-spin density matrix, see Eq.~\eqref{eqHFRDM1}. The comparison in Fig.~\ref{figAFM_FCS} (a) yields an estimate $c_{p \gtrsim 0.1} \approx 0.07$.

In Fig.~\ref{figAFMFiniteHoleDoping} (b) we also show the AFM order parameter $\mean{S_z^i}$, which takes non-zero values only if the $\rm SU(2)$-symmetry is spontaneously broken. For all temperatures, we observe a transition from a phase with broken ${\rm SU(2)}$ symmetry at low doping, to an ${\rm SU(2)}$-symmetric phase at higher doping. The transition point shifts to higher doping values when the temperature is decreased. As a result of quantum fluctuations, the non-collinear correlations $\mean{S_i^{+}S_j^{-}}$ develop when the doping is increased, and the collinear correlations $\mean{S_i^{z}S_j^{z}}$ are strongly reduced compared to their value $\mean{S_i^{z}S_j^{z}}_{p=0,T=0} = -0.25$ in the classical N\'eel state. The latter is obtained as the mean-field solution at half filling and zero temperature.

Indeed, the ${\rm SU}(2)$ symmetry breaking phase transition is clearly visible in the FCS of the system-averaged two-spin density matrix. In Fig. \ref{figAFM_FCS} (a) we show the FCS of the system-averaged triplet probability, $p_t = \bra{t} \rho^S \ket{t}$ at a fixed temperature $T=0.125J$ and for various doping values, ranging between $p=0.01$ to $p=0.05$. We observe how the SU(2) symmetry is gradually restored and the distribution function narrows when the critical doping value, where the transition takes place, is approached. As demonstrated in Fig.~\ref{figAFM_FCS} (b), increasing the temperature at a fixed doping value has a similar effect on the FCS. Beyond the critical doping, respectively temperature, a sharp peak remains and the ${\rm SU}(2)$ symmetry is fully restored. For a fully mixed state at infinite temperature the triplet probability is $p_t^{T \rightarrow \infty} = 0.25$, slightly higher than the value $p_t \approx 0.23$ which we predict in the $\rm{SU}(2)$ symmetric phase at large doping values.


\begin{figure*}[t!]
\centering\
\epsfig{file=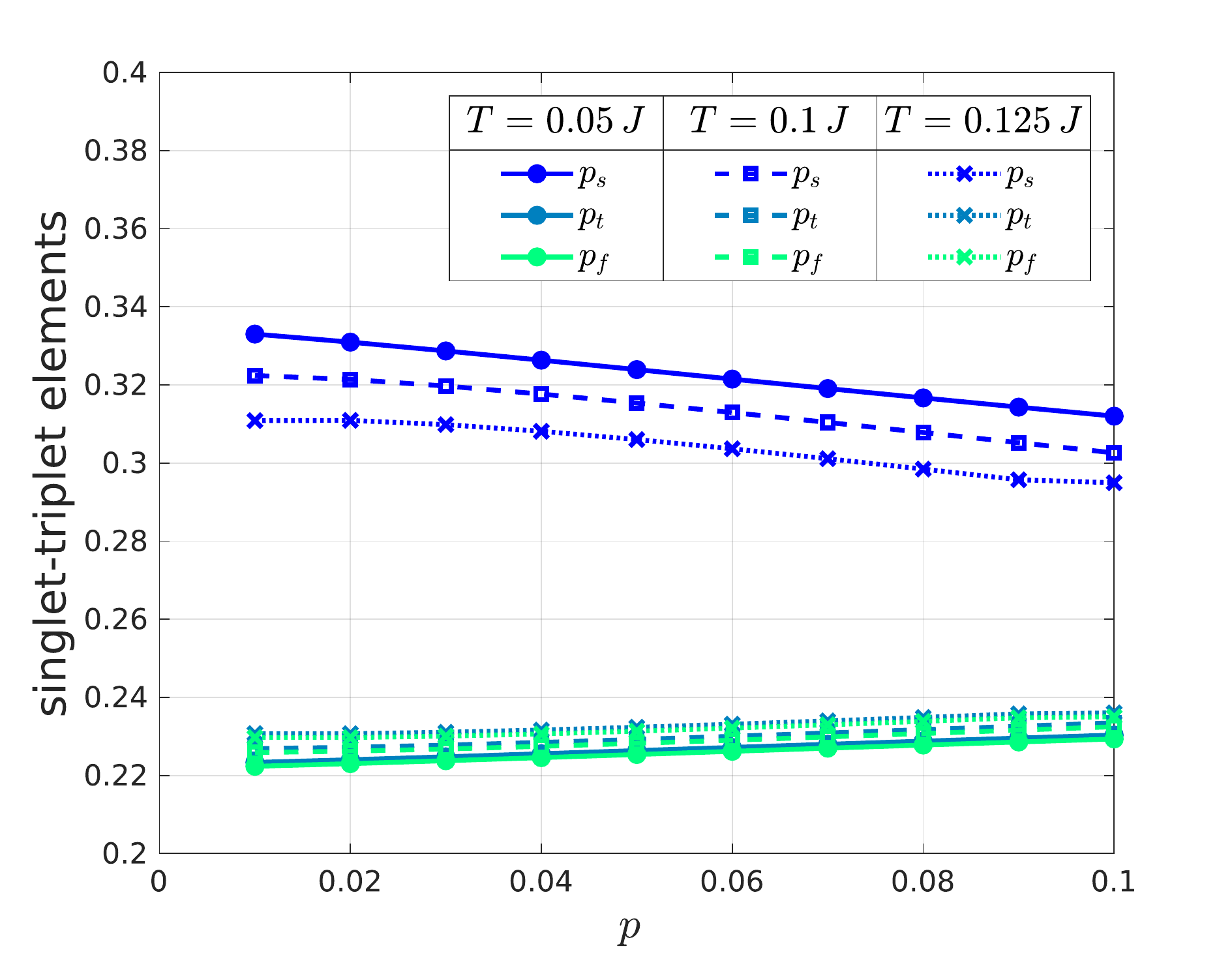, width=0.48\textwidth}
\epsfig{file=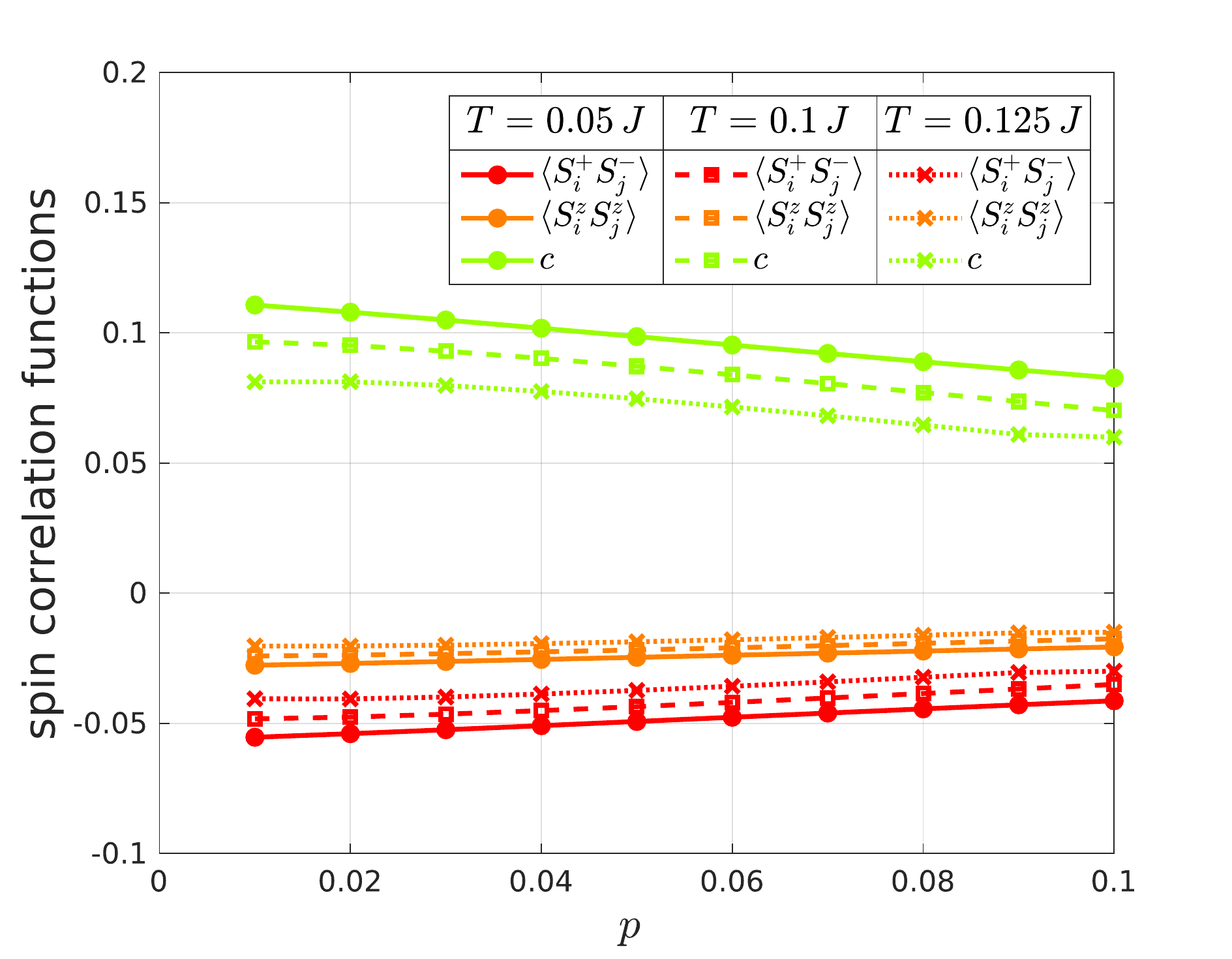, width=0.48\textwidth}
\caption{Doping-dependence of local correlations in the FL* phase. We show the doping dependence of the short-range singlet, triplet and ferromagnetic amplitudes ($p_s$, $p_t$ and $p_f$) in (a), for various temperatures ($T = 0.05 J$ (bold line), $T = 0.1J$ (bold dashed) and $T=0.125J$ (dotted)) at $t=4J$. In (b) the doping dependence of spin correlation functions and the parameter $c$, as defined in Eq.~\eqref{eqHFRDM1}, is shown for the same parameters. The singlet amplitude in (a) decreases for larger doping and temperature due to the presence of holes, respectively thermal fluctuations. Since the state is SU(2) symmetric, triplet and ferromagnetic amplitude are identical. The magnitude of short-range spin correlations decreases by the presence of holes that destroy the AFM background ordering.} 
\label{figSLFiniteHoleDopingC}
\end{figure*}

\subsection{Fractionalized Fermi liquid}

As a second example we consider signatures of a FL* phase in the reduced density matrix. This phase is SU(2) symmetric and the mean-field amplitudes $n_z$ and $m_z$ in Eq. \eqref{MF_mz}-\eqref{MF_nz} vanish identically.
 Only the mean-fields $\chi_{ij}$ and $\Delta_{ij}$ remain finite. Importantly, there is no hybridization between dopons and spinons, i.e.~$\langle f^\dagger_i d_i \rangle = 0$. The localized spins are thus in a spin-liquid state and the dopons form a Fermi liquid with a small Fermi surface ($\sim p$) on the spin-liquid background. Note that a finite hybridization gives rise to a "heavy Fermi liquid" in the Kondo-Heisenberg terminology, which corresponds to an ordinary Fermi liquid phase in the corresponding $t-J$ model and is expected to appear only at large hole doping levels (see Ref. \cite{Ribeiro2006} for a detailed discussion).

After solving the self-consistency equations under the condition $n_z=m_z=0$ (see appendix  \ref{AppendixDopedCarrier}), we determine the reduced density matrix in the singlet and triplet basis. The results are summarized in Fig. \ref{figSLFiniteHoleDopingC}. Due to the presence of holes, the singlet amplitude $p_s$ of nearest neighbor spins, as shown in part (a) of this figure, slowly decreases as function of doping. Thermal fluctuations reduce the singlet amplitude further, but the qualitative doping dependence of the curves remains independent of temperature. Because the state is ${\rm SU}(2)$ symmetric, the triplet amplitude $p_t = p_f$ is equal to the probability to find ferromagnetically aligned spins. This can be seen by noting that $p_{t}-p_{f} = \mean{S_i^x S_j^x}+\mean{S_i^y S_j^y}-2\mean{S_i^z S_j^z} = 0$. Both $p_t=p_f$ increase when the singlet amplitude $p_s$ decreases. The singlet-triplet matrix element $\langle s | \rho^S | t \rangle = 0$ vanishes because of the ${\rm SU}(2)$ symmetry, and is not shown in the figure.

In Fig.~\ref{figSLFiniteHoleDopingC} (b) we calculate the doping dependence of spin-spin correlations, for which $\mean{S_i^z S_j^z} = \mean{S_i^{+}S_j^{-}} /2$ because of the ${\rm SU}(2)$ symmetry. On nearest neighbors they are very weakly doping dependent and remain negative, corresponding to weak and short-range AFM correlations in the system. The $\rm{SU}(2)$ invariant two-spin density matrix can be characterized by the parameter $c$, see Eq.~\eqref{eqHFRDM1}, which starts at $c_{p=0.01} \approx 0.08$ for very small doping and continuously decreases to $c_{p=0.1} \approx 0.06$ at higher doping.

\subsection{Numerical results for the $t-J$ model}

Next we perform a numerical study of the two-spin reduced density matrix $\rho^{\rm S}$ in a periodic $4 \times 4$ lattice with one hole; this corresponds to a doping level of $p \approx 6 \%$. We perform exact diagonalization (ED) to calculate the zero-temperature ground state, in a sector of the many-body Hilbert space where the single hole carries total momentum $\vec{k}=(\pi/2,\pi/2)$ and the total spin in $z$-direction is $S^z=1/2$. This state describes a magnetic polaron, the quasiparticle formed by a single hole moving in an AFM background \cite{schmitt1988spectral,kane1989motion,sachdev1989hole,liu1992dynamical,brunner2000single,white2001density}. Even though the considered system size is small, we expected that the local correlations encoded in the two-spin density matrix are close to their values in an infinite system at $6 \%$ doping. Because of the limited size of the lattice, we refrain from calculating the FCS of the system-averaged two-spin density matrix. 

To study the effect that the mobile hole has on the surrounding spins, we tune the ratio $J/t$. Although not identical, we expect that the effects of larger tunnelings $t/J$ and higher doping $p$ are comparable in the finite-size system: When $t \gg J$, the hole is moving faster through the anti-ferromagnet, thus affecting more spins. Indeed, when $J / t \gg 1$ the hole is quasi-static and the surrounding spins have strong AFM correlations; on neighboring sites their strength approaches their thermodynamic values in the two-dimensional Heisenberg model. 

\begin{figure}[b!]
\centering
\epsfig{file=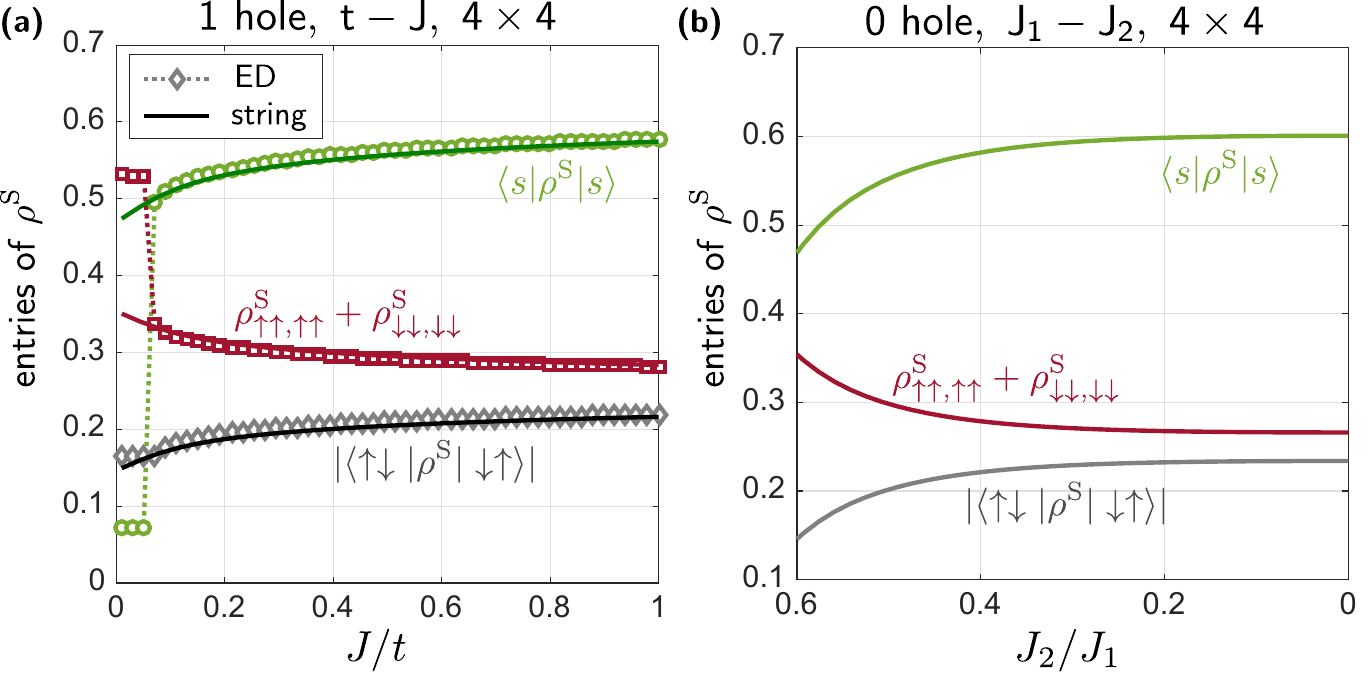, width=0.5\textwidth}
\caption{(a) Entries of the reduced two-spin density matrix in a $4 \times 4$ system doped with a single hole, as a function of $J/t$. We used ED simulations for the ground state with total momentum $\vec{k}=(\pi/2,\pi/2)$ and with total spin $S^z=1/2$. We compare our numerically exact results (symbols) to predictions by the geometric string theory (solid lines), for details see Ref. \cite{grusdt2018parton}, where the ratio $J/t$ tunes the length of the string of displaced spins. The reduced two-spin density matrix of the doped system closely resembles the two-spin density matrix in an un-doped but frustrated quantum magnet: In (b) we calculate $\rho^{\rm S}$ for a frustrated $J_1-J_2$ model on a periodic $4 \times 4$ lattice with $S^z=0$, with diagonal next-nearest neighbor couplings $J_2$. Locally the two systems become indistinguishable, and we find that larger values of $t/J$ correspond to larger values of $J_2 / J_1$.} 
\label{figSingleHoleDoping}
\end{figure}

In Fig.~\ref{figSingleHoleDoping} (a) we show how the entries of the nearest-neighbor two-spin density matrix depend on the ratio $J/t$. When $J / t =1$ we find that the singlet probability $p_{\rm s}$ in the doped system is still close to the value $0.57$ expected from linear spin-wave theory in an un-doped system. For smaller values of $J/t$ the hole has a more pronounced effect on the spin environment, which leads to a decrease of the singlet probability $p_{\rm s}$. In addition, the probability $p_{\rm f}$ to find ferromagnetic correlations increases to values larger than $0.29$ expected at zero doping from linear spin-wave theory. For very small values of $J/t < 0.1$, we observe a phase transition in the finite-size system, which is expected to be related to the formation of a Nagaoka polaron \cite{nagaoka1966ferromagnetism,white2001density}.

Qualitatively similar behavior is expected in an undoped system when frustrating next-nearest neighbor couplings $J_2$ are switched on, in addition to the nearest-neighbor interactions $J_1=J$. To demonstrate this, we use ED to calculate the reduced two-spin density matrix in a $J_1-J_2$ model on the same $4 \times 4$ lattice. As shown in Fig.~\ref{figSingleHoleDoping}, increasing $J_2$ from zero to $J_2  = 0.5 J_1$ has a similar effect as decreasing $J/t$ from a value of $1$ to $0.1$. I.e., locally the doped $t-J$ model cannot be distinguished from a frustrated quantum magnet described by the $J_1-J_2$ Hamiltonian. 

Our exact numerical results in the $4\times4$ system are consistent with the physical picture derived previously from the doped carrier formalism. Smaller values of $t/J$, expected to reflect high doping values, lead to a decrease of the singlet amplitude, which is directly related to the $\mean{\vec{S}_i \cdot \vec{S}_j}$ correlations, on nearest neighbor sites. 

Additional quantitative understanding of the $J/t$-dependence in the single-hole problem can be obtained by the geometric string approach introduced in Refs.~\cite{grusdt2018parton,grusdt2018meson}. There, one describes the motion of the hole along a fluctuating string of displaced spins and applies the frozen-spin approximation \cite{grusdt2018meson}: It is assumed that the quantum state of the surrounding spins is determined by a parent state $\ket{\tilde{\Psi}}$ in the undoped system, and the hole motion only modifies the positions of the parent spins in the two-dimensional lattice, otherwise keeping their quantum states unmodified. To calculate the two-spin density matrix $\rho^S_{\rm NN}$ on a given nearest-neighbor bond, we trace over all possible string configurations. Because the strings modify the positions of the parent spins, $\rho^S_{\rm NN}$ describes a statistical mixture of nearest neighbor ($\tilde{\rho}^S_{\rm NN}$), next-nearest neighbor ($\tilde{\rho}^S_{\rm NNN}$),... two-spin density matrices, with coefficients $p_{\rm NN}$, $p_{\rm NNN}$,... . The results in Fig.~\ref{figSingleHoleDoping} (a) (solid lines) are obtained by using the exact ground state of the undoped Heisenberg model in the $4 \times 4$ lattice as the parent state. The weights $p_{\rm NN}$, $p_{\rm NNN}$,... are determined by averaging over string states with a string length distribution calculated as described in Ref. \cite{grusdt2018parton}.

\section{Conclusion and Outlook}

Our work demonstrates that a magnetically ordered state can be identified by measuring the statistical distribution of the nearest-neighbor triplet amplitude of the system-averaged two-spin density matrix, which arises due to random orientations of the order parameter between different experimental shots. In fact, it is sufficient to measure the FCS of a generic local operator which does not transform like an SU(2) singlet in order to identify the AFM phase from local measurements. Moreover, we have calculated the nearest neighbour singlet and triplet amplitudes as a function of the hole concentration away from half filling within the doped carrier formalism and demonstrated that the triplet probability distribution has a finite width in the magnetically ordered phase, which decreases continuously with doping and temperature. At the phase transition from the magnetically ordered to a paramagnetic state, such as the FL*, the distribution turns into a sharp peak.

The fact that the information about symmetry broken states is contained in the FCS of the system-averaged two-spin density matrix shows that the FCS distribution can be measured experimentally without the use of a quantum gas microscope. Experiments with superlattice potentials where the average over all double-wells is taken automatically during the readout after each shot work equally well. This might even be an advantage due to the larger system sizes that can be reached compared to setups with a quantum gas microscope.

Even though our work focused on AFM ordered states, we emphasize that measuring the FCS with respect to different order parameter realizations can also be used to detect other types of broken symmetries, such as states that break lattice symmetries like charge-density waves or valence bond solids. We also note that our analysis can be straightforwardly extended to study correlations beyond nearest neighbors.

The tools introduced in this work potentially allow quantum gas microscopy experiments at currently accessible temperatures to shed light on the long-standing puzzle about the nature of the pseudogap state in underdoped cuprate superconductors. Another interesting route to study effects of doping a Mott insulator is to measure the single particle spectral function in analogy to ARPES experiments in the solid-state context \cite{Bohrdt2018}. In combination with the tools discussed in this work, quantum gas microscopy should be able to characterize the properties of doped Mott insulators to a high degree, providing a valuable benchmark for theoretical proposals.

\acknowledgements

We thank E. Demler, M. Kanasz-Nagy, Richard Schmidt, D. Pimenov, Annabelle Bohrdt and Daniel Greif for valuable discussions. S. H. and M. P. were supported by the German Excellence Initiative via the Nanosystems Initiative Munich (NIM). F. G. acknowledges financial support by the Gordon and Betty Moore foundation under the EPiQS program.

\appendix\label{Appendix}

\begin{figure*}[ht!]
\centering\
\epsfig{file=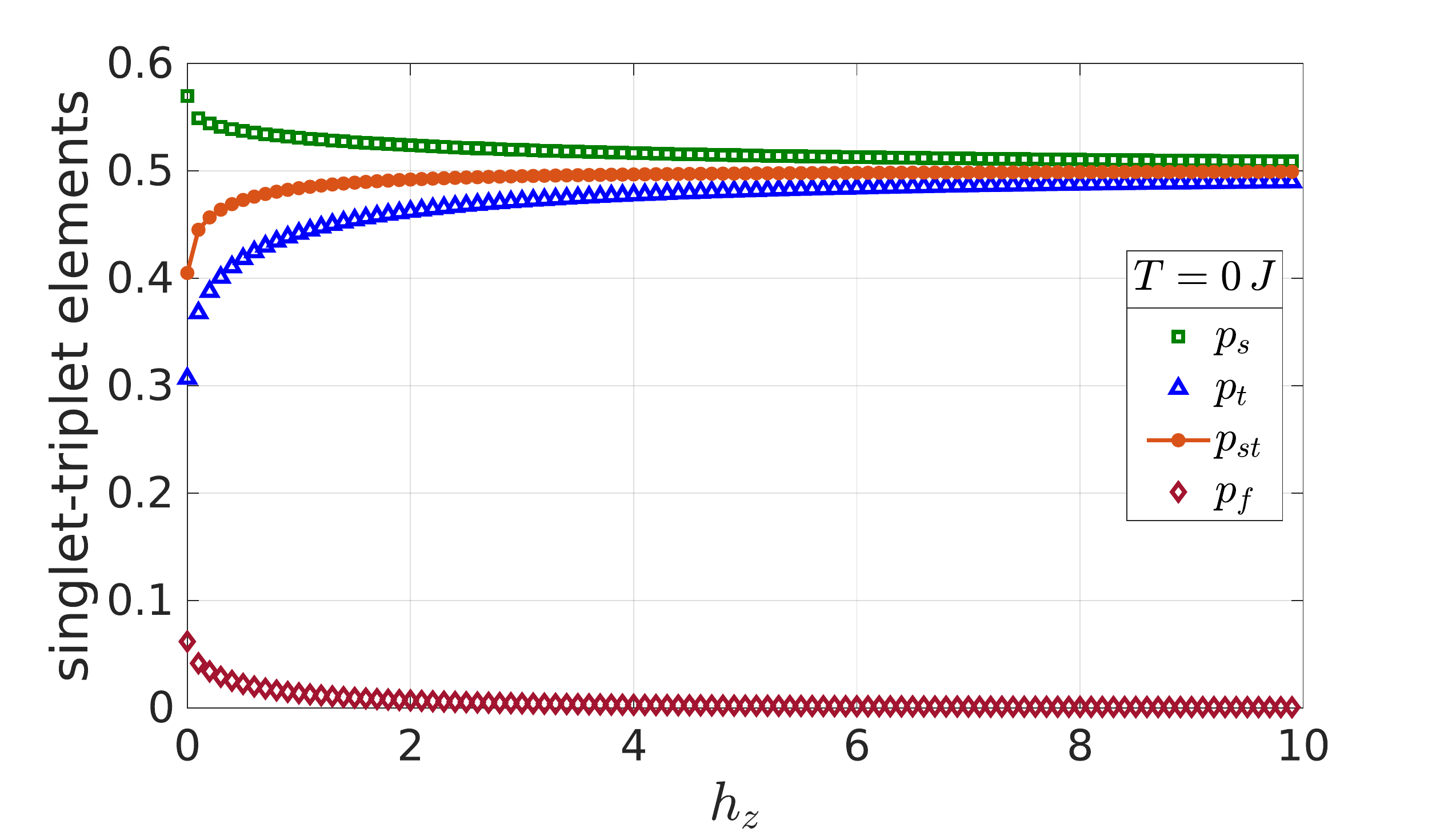, width=0.5\textwidth}
\epsfig{file=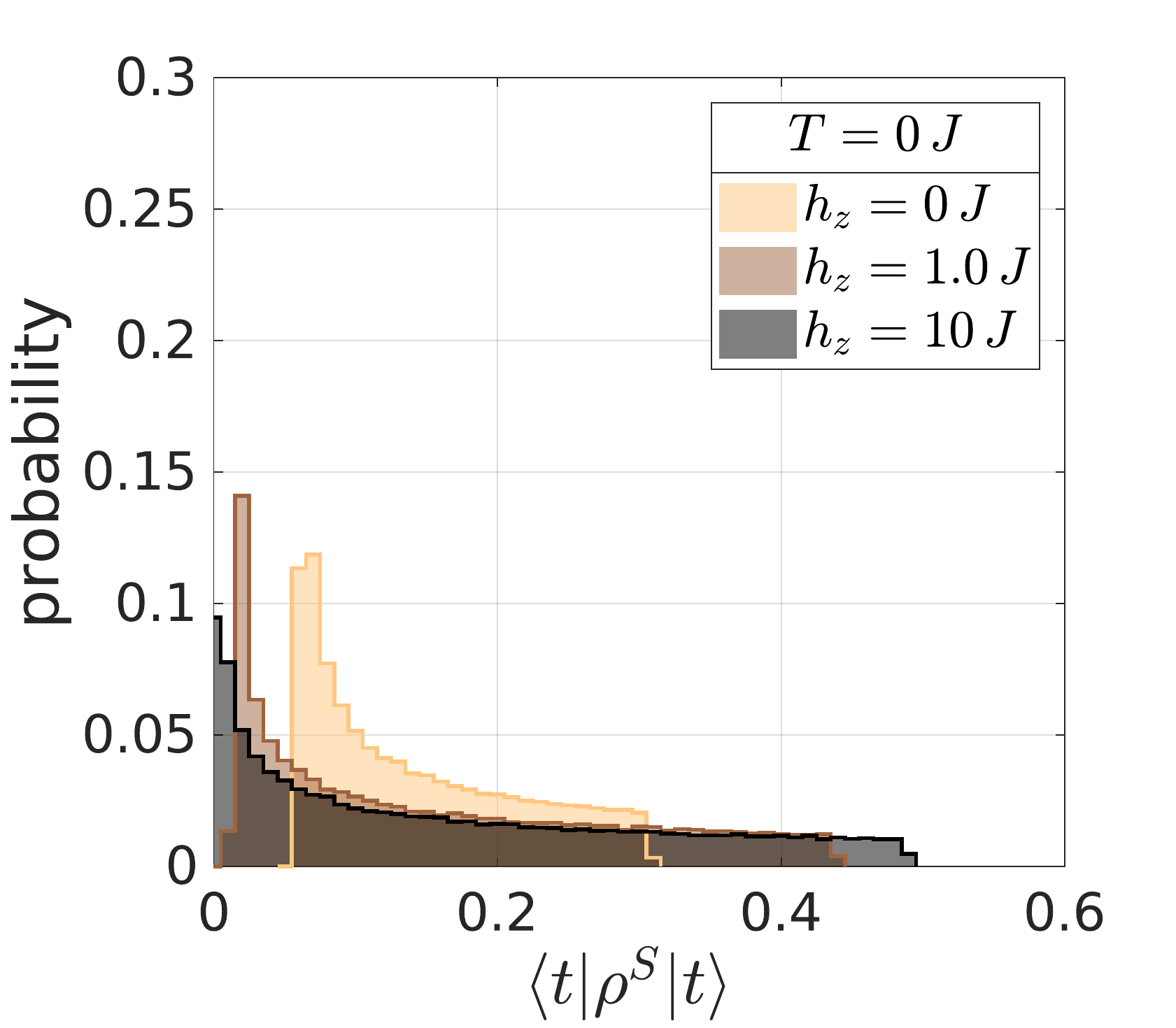, width=0.334\textwidth}
\caption{Part (a) shows the singlet $p_{\text{s}}$, triplet $p_{\text{t}}$ and ferromagnetic $p_{\text{f}}$ matrix elements as a function of the external staggered field $h_z$. 
Increasing the staggered field drives the state closer to a classical N\'{e}el configuration. 
Part (b) shows the full counting statistics of the triplet amplitude for different values of $h_z$. A strong external staggered magnetic field suppress quantum fluctuations and we observe a continuous distribution of the triplet amplitude in the range $0 \leq \bra{t} \rho^S \ket{t} \leq 0.5$. Since quantum fluctuation locally disturb the antiparallel alignement, the distribution has an onset at a finite value for a small external staggered fields.} 
\label{figSpinWave}
\end{figure*}

\section{Symmetry of the reduced density matrix}\label{AppendixSymRed}

The block diagonal form of the reduced density matrix in section \ref{SecFingerprintsQSL} is a direct consequence of the global particle conservation. This is actually shown for a pure state in Ref. \cite{li2008entanglement} using a singular value decomposition. In the following we verify that this also holds for any thermalized system with a global conserved quantity under certain, rather weak, limitations.

We thus start with a Hamiltonian with a conserved quantity $\mathcal{O}$, i.e. $[H,\mathcal{O}]=0$, and an associated system that is described by the density matrix $\rho =  e^{-\beta H}$. Quite generally, the operator $\mathcal{O}$ can be decomposed as $\mathcal{O} = \sum_{j} \mathcal{O}_{j}$, where $j$ labels e.g.~different lattice sites. In the absence of spontaneous symmetry breaking the operator $\mathcal{O}$ also commutes with the density matrix $[\rho,\mathcal{O}]=0$, i.e.~the state $\rho$ has the same symmetry as the Hamiltonian. In such a case we can show that the reduced density matrix $\rho_{A}=\text{tr}_{\bar{A}}(\rho)$ of a subsystem $A$ commutes with the operator $\mathcal{O}_{A} = \sum_{j \in A} \mathcal{O}_{j}$ (here $\bar{A}$ is the complement of $A$):

\begin{align}
[\rho_A , \mathcal{O}_{A}] &= \text{tr}_{\bar{A}}(\rho) \mathcal{O}_{A} - \mathcal{O}_{A} \text{tr}_{\bar{A}}(\rho) = \text{tr}_{\bar{A}}(\rho \mathcal{O}_{A} - \mathcal{O}_{A} \rho),\\
0 &= \text{tr}_{\bar{A}}([\rho, \mathcal{O}]) =  \text{tr}_{\bar{A}}(\rho \mathcal{O}_{A} - \mathcal{O}_{A} \rho) + \text{tr}_{\bar{A}}(\rho \mathcal{O}_{\bar{A}} - \mathcal{O}_{\bar{A}} \rho).
\end{align}

If we now use that the trace is cyclic, i.e. $\text{tr}_{\bar{A}}(\rho \mathcal{O}_{\bar{A}}) = \text{tr}_{\bar{A}}(\mathcal{O}_{\bar{A}} \rho)$, and combine this with Eq. (1) and (2) from above, we see immediately that $[\rho_A , \mathcal{O}_{A}] = 0$.  So we can conclude that the two-site density matrix can always be written in a block diagonal form, if the above requirements are satisfied. In section \ref{Sec2SpRedDenMat} of the main text we use this result to show that the global U(1) symmetry (i.e.~particle number conservation) of the Fermi-Hubbard model together with the fact that we only consider states with a definite particle number implies that the reduced density matrix can be written in a block diagonal form, where each block can be labeled by the number of electrons in the subsystem. This is independent of the presence or absence of long-range magnetic order, which only has consequences for the block diagonal form of the reduced density matrix in different spin sectors, but does not affect the block diagonal form in the different particle number sectors.

\section{Spin wave theory at half filling}\label{AppendixHolPrim}

In this first part of the appendix we want to summarize a Holstein-Primakoff analysis (HP) of the antiferromagnetic Heisenberg model on a bipartite lattice at half filling \cite{Auerbach2012}. We thus aim to study the spin system deep in an AFM phase, where neighboring spins tend to point in opposite directions. In order to tune the strength of quantum fluctuations, we allow for an additional external staggered magnetic field $h_z$ along the $z$-direction, which explicitly breaks the SU(2) invariance of the system and fixes the magnetization direction. We associate all spins pointing upwards (downwards) with sublattice $A$ ($B$). Quantum fluctuations around the classical N\'{e}el state are represented by bosonic excitations in the HP analysis.

\textit{Method} - This quite standard approach is based on a canonical mapping between spin and bosonic operators given by
\begin{align}
S^{z}_i &= ( S - \cre{b}{i} \ann{b}{i} ), & S^{z}_j &= (- S + \cre{b}{j} \ann{b}{j} ), \\
S^{-}_i &\simeq \sqrt{2S} \cre{b}{i}, & S^{+}_j &\simeq \sqrt{2S} \cre{b}{j},\\
S^{+}_i &\simeq \sqrt{2S} \ann{b}{i}, & S^{-}_j &\simeq \sqrt{2S} \ann{b}{i},
\end{align}
where $i \in A$ and $j \in B$ and we have taken the semi-classical large $S$ limit. Furthermore, we have to constrict the local Hilbert space by $2n_{b,i} \leq S$, where $n_{b,i}$ is the boson occupationj on site $i$. We now perform a rotation around the x-axis on sublattice $B$ and expand the Heisenberg model in $1/S$. For a small number of excitations, i.e.~$|S^{z}_i | \approx S$, we can neglect terms of order $\mathcal{O}(1/S)$ and the spin wave Hamiltonian takes the form
\begin{align}
H^{SW} =& - S^2 J N \frac{z}{2} - S J N \frac{z}{2} + \sum_{\textbf{k}} \omega_{ \textbf{k} } [ \cre{a}{ \textbf{k} } \ann{a}{ \textbf{k} } + \frac{1}{2} ],\label{SWH}
\end{align}
where $z$ is the coordination number and
\begin{align}
\gamma_{ \textbf{k} } &= \tfrac{1}{2} [ \cos{ k_x } +\cos{ k_y } ],\\
\ann{a}{\textbf{k}} &= \cosh{ \theta_{\textbf{k}} } \ann{b}{\textbf{k}} - \sinh{ \theta_{\textbf{k}} } \cre{b}{\textbf{-k}},\\
\tanh{( 2 \theta_{\textbf{k}} )} &= - \Delta \gamma_{ \textbf{k} } \text{ with } \Delta = \frac{JSz}{JSz-h_z},\\
\omega_{ \textbf{k} } &= \vert J \vert S z \sqrt{1-\gamma_{ \textbf{k} }^2}.
\end{align}
The ground state has no magnon excitation, i.e. $\ann{a}{\textbf{k}}\ket{\text{GS}} = 0$ and the ground state energy is then $ E_0 = - J N \frac{z}{2} S(S+1) + \frac{1}{2} \sum_{\textbf{k}} \omega_{ \textbf{k} }$. The ground state wavefunction itself has the form
\begin{align}
\ket{\text{GS}} = \mathcal{N} \exp{[ \frac{1}{2} \sum_{ \textbf{k} } \tanh{ \theta_{\textbf{k}} } \cre{b}{\textbf{k}} \cre{b}{\textbf{-k}} ]} \ket{0},\label{GSAF}
\end{align}
where $\mathcal{N}$ is a normalization constant. 

\emph{Results.--}
We now determine the two-particle reduced density matrix at zero temperature by computing the expectation values $\mean{ S^{\alpha}_i S^{\beta}_j }$ (see Eq.~\eqref{rhoS1}) for the AFM ground state wavefunction at zero temperature. The entries of the reduced density matrix show that the system is close to a N\'{e}el state, so that the matrix element in Eq. \eqref{eqDefpst} is given by
\begin{align}
p_{\text{st}} &= M_z,\label{psinglettripletAF}
\end{align}
with the local staggered magnetization
\begin{align}
M_z &= |\mean{S^{z}_i}| = S - \tfrac{1}{4 \pi^2} \int_{\text{1st BZ}}d^2 k \sinh^2{(\theta_{\textbf{k}})}.
\end{align}
When the external staggered field is absent, i.e.~$h_z = 0$, the local staggered magnetization smaller than the total spin $S$ due to quantum fluctuations. Increasing the external staggered field, the excitation of bosonic quasiparticles gets more unfavorable, so that the local magnetization converges against the total spin $M_z \rightarrow S$. This is also shown Fig.~\ref{figSpinWave}(a), which shows the convergence of $p_{st}$ against the total spin $S$ with increasing $h_z$. The difference of singlet and triplet elements is given by 
\begin{align}
p_{\text{s}} - p_{\text{t}} &= 2g_z =- [\mean{S^{-}_i S^{+}_j}+\mean{S^{+}_i S^{-}_j}] \label{diffpsingletptripletAF} \\
&= -\tfrac{1}{4 \pi^2} \int_{\text{1st BZ}}d^2 k \cos{(\bold{k} \bold{d})}\sinh{(2\theta_{\textbf{k}})}, \nonumber
\end{align}
with $\bold{d} = \bold{R}_i - \bold{R}_j$. It decreases from approximately $g_z(h_z=0) \approx 0.13$ to $g_z(h_z=10) \approx 0.01$ (see Fig. \ref{figSpinWave} (a)) and thus shows that the state associated with the reduced density matrix approaches an equal superposition of triplet and singlet component when increasing the external staggered magnetic field $h_z$. This is in agreement with the former result that the three elements $p_{\text{st}}, p_{\text{t}}$ and $p_{\text{s}}$ converge to the total spin $S$ with growing $h_z$. Finally the ferromagnetic element decreases with increasing $h_z$ and is given by (see Eq.~\eqref{eqDefpf} )
\begin{align}
p_{\text{f}} =\tfrac{1}{4}-M_z^2-g_z^2-f_z^2, \label{pferroAF}
\end{align}
where
\begin{align}
f_z = \tfrac{1}{4 \pi^2} \int_{\text{1st BZ}}d^2 k \cos{(\bold{k} \bold{d})}\sinh^2{(\theta_{\textbf{k}})}. 
\end{align}
By explicitly breaking the SU(2) invariance due to the external staggered magnetic field any ferromagnetic alignment of neighbouring spins is unfavorable. We observe this effect also in the FCS shown in Fig. \ref{figSpinWave}(b). For further discussion see also main text section \ref{SecSigFHM}.

\section{Doped carrier analysis of the $t-J$ Hamiltonian}\label{AppendixDopedCarrier}

Here we complete the doped carrier mean-field calculation introduced in the main text in section \ref{SecSigBHF} based on Ref.~\cite{Ribeiro2006}. 
Ribeiro and Wen showed that the phase diagram generally includes a pseudogap regime, where the mean-field amplitudes are in agreement with a fully SU(2) symmetric ground state for a small parameter regime of doping $0 \lesssim p \lesssim 0.1$ and temperature $0.1J \lesssim T \lesssim 0.2J$ consistent with a pseudogap metal, where spinon and dopon do not hybridise and no ordering developes in the system (see phase diagram in Ref.~\cite{Ribeiro2006}). In this phase fermionic spinons form d-wave Cooper pairs and a gap opens in the electronic spectral function, as expected from ARPES measurements in the pseudogap regime of underdoped cuprates. Moreover, the phase diagram has a very dominant antiferromagnetic phase centered around half filling, which gets more pronounced when the ratio $t/J$ is lowered. Other phases such as a d-wave superconductor and an ordinary metal are also present in the phase diagram. These phases are characterized by a finite hybridisation matrix element between spinons and dopons.

In the following we show the mean-field Hamiltonian in momentum space and determine the self-consistency equations. We restrict our analysis on special regimes of the mean-field phase diagram. As discussed in more detail in Sec.~\ref{SecSigBHF} of the main text we are interested in the antiferromagnetic phase at low doping, as well as in the pseudogap regime, which is modelled as a fractionalized Fermi liquid in this approach. 

\emph{Overview on mean-field amplitudes.--}
Let us first assume that the system exhibits a finite magnetization $\mean{\bold{S}_i} \neq 0$. We fix the magnetization direction to be along the z-axis and define the following operators
\begin{align}
m_i^z \sigma^z_{\alpha \beta} &= \cre{f}{i,\alpha} \ann{f}{i,\beta} ,\\
n_i^z \sigma^z_{\alpha \beta} &= -\frac{1}{4} \sum_{\nu \in \{ 2,3\} } \frac{t_{\nu}}{8} \sum_{\hat{u}_v} \cre{d}{i,\alpha} \ann{d}{i+\hat{u}_v,\beta},
\end{align}
where the latter includes further neighbor hopping amplitudes $t_v$ with $v \in \{ 2,3 \}$ and the second sum runs over the respective unit vectors $\hat{u}_2 = \pm e_{\hat{x} / \hat{y}} \pm e_{\hat{y} / \hat{x}}$ and $\hat{u}_3 = \pm  2e_{\hat{x} / \hat{y}}$. We consider then the following mean-field amplitudes $m^z = (-1)^{i_x + i_y} \mean{m_i^z}$ and $n^z = (-1)^{i_x + i_y}\mean{n_i^z}$. The amplitude $n^z$ measures thereby the hopping of dopons and describes a net magnetization of holes with respect to an AFM ordered spin background.

In principle it is also possible that spinons and dopons hybridise, although in regimes of our interest this is not the case, we consider them here for completeness. If the ground state is for instance a normal Fermi liquid, we expect this amplitude to be finite. The hybridization between spinons and dopons is described by the operator $\ann{\kappa}{ij}\delta_{\alpha\beta}=2\cre{d}{i,\alpha}\ann{f}{j,\beta}$ and its expectation values
\begin{align}
b_0 &= \mean{\kappa_{ii}},\label{MF_b0}\\
b_1 &= \tfrac{3}{16}\sum_{\nu \in \{ 2,3\} } \frac{t_{\nu}}{8} \sum_{j=i+\hat{u}_v} \mean{\kappa}_{ij}.\label{MF_b1}
\end{align} 
Finally we also recap that the spinons are assumed to form Cooper-pairs in the d-wave channel $\Delta$ and we also allow for a finite spinon hopping amplitude $\chi$ as described in section \ref{SecSigBHF}.

\emph{Mean-field Hamiltonian.--}
After decoupling the Hamiltonian in these channels and using the compact Nambu notation $\cre{\psi}{k}=[\cre{f}{k,\uparrow},\ann{f}{-k,\downarrow}]$ and $\cre{\eta}{k}=[\cre{d}{k,\uparrow},\ann{d}{-k,\downarrow}]$, the mean-field Hamiltonian in momentum space takes the form
\begin{widetext}
\begin{align}
H^{\text{MF}}_{tJ}=&\tfrac{3}{4}\tilde{J}N(\Delta^2+\chi^2)-N\mu_d+2\tilde{J}N|m^z|^2-4 N n^z m^z-4Nb_0 b_1-N\mu_d \label{MFHamiltonian}\\
&+\sum_{\bold{k}}[\cre{\psi}{k}\ \cre{\eta}{k}]
\begin{bmatrix}
\alpha^x_k \sigma_x+\alpha^z_k \sigma_z & \beta_k \sigma_z\\
\beta_k \sigma_z & \gamma_k \sigma_z
\end{bmatrix}
\begin{bmatrix}
\ann{\psi}{k}\\
\ann{\eta}{k}
\end{bmatrix}
+\sum_{\bold{k}}[\cre{\psi}{k+Q}\ \cre{\eta}{k+Q}]
\begin{bmatrix}
[4 n^z - 4 \tilde{J} m^z] \sigma_0 & 0\\
0 & -4 m^z[\gamma_k + \mu_d] \sigma_0
\end{bmatrix}
\begin{bmatrix}
\ann{\psi}{k}\\
\ann{\eta}{k}
\end{bmatrix}\nonumber
\end{align}
\text{where}\\
\begin{align}
\alpha^x_k &=-\tfrac{3}{4}\tilde{J}\Delta[\cos{(k_x)}-\cos{(k_y)}],\\
\alpha^z_k &=\lambda+p t_1[\cos{(k_x)}+\cos{(k_y)}]-\tfrac{3}{4}\tilde{J}\chi[\cos{(k_x)}+\cos{(k_y)}],\\
\beta_k &=b_1+\tfrac{3}{8}b_0[t_1[\cos{(k_x)}+\cos{(k_y)}]+2t_2[\cos{(k_x)}\cos{(k_y)}]+t_3[\cos{(2k_x)}+\cos{(2k_y)}]],\\
\gamma_k &= \tfrac{1}{2}t_2\cos{(2k_x)}\cos{(2k_y)}+\tfrac{1}{4}t_3[\cos{(2k_x)}+\cos{(2k_y)}]-\mu_d.
\end{align}\label{MFH_MS}
\end{widetext}
The chemical potential $\mu_d$ is used to adjust the dopon density in the system. We further use $\sigma_0=\mathbbm{1}_{2 \times 2}$ and $Q=(\pi/a,\pi/a)$.

\emph{Antiferromagnetic metal: $b_0=b_1 = 0; n^z \neq 0, m^z \neq 0$.--} 
The hybridisation between spinons and dopons of the form $\langle  f^\dagger_i d_i \rangle$ is now neglected, i.e. $b_0 = b_1 \approx 0$, since magnetic ordering and superconductivity are not assumed to be present at the same time. In this case the spinon-density Lagrange parameter $\lambda=0$. The eigenenergies of the above mean-field Hamiltonian $H^{\text{MF}}_{tJ}$ for the spinon and dopon sector then read
\begin{align}
\epsilon^{\pm}_{s,\bold{k}} &= \pm \sqrt{(\alpha_{k}^x)^2 + (\alpha_{k}^z)^2 + (\nu_{k})^2},\\
\epsilon^{\pm}_{d,\bold{k}} &= (1 \mp 4 |m^z|) -\mu_d,
\end{align}
where $\nu_k = -4(\tilde{J}m^z - n^z)$. The set of self-consistency equations are determined from minimizing the free energy density $f=F/N$ for a fixed density of dopons $p$ and read
\begin{align}
&m^z  - \frac{1}{N} \sum_k \frac{\tilde{J}m_s^z-n^z}{\epsilon_{s,k}^{+}}\frac{\sinh{(\beta\epsilon_{s,k}^{+})}}{1+\cosh{(\beta\epsilon_{s,k}^{+})}}= 0,\label{SCE_M_1}\\
&n^z  + \frac{1}{2N} \sum_k \gamma_k \frac{\sinh{(\beta (\gamma_k -\mu_d))}}{1+\cosh{(\beta(\gamma_k -\mu_d))}} = 0,\nonumber\\
&\chi  + \frac{1}{2N} \sum_k \frac{\alpha_k^z [\cos{(kx)}+\cos{(ky)}]}{\epsilon_{s,k}^{+}}\frac{\sinh{(\beta\epsilon_{s,k}^{+})}}{1+\cosh{(\beta\epsilon_{s,k}^{+})}} = 0,\nonumber
\end{align}
\vspace{0.04cm}
\begin{align}
&\Delta  + \frac{1}{2N} \sum_k \frac{\alpha_k^x [\cos{(kx)}-\cos{(ky)}]}{\epsilon_{s,k}^{+}}\frac{\sinh{(\beta\epsilon_{s,k}^{+})}}{1+\cosh{(\beta\epsilon_{s,k}^{+})}} = 0,\nonumber\\
&x - 1 + \frac{1}{N} \sum_k \frac{\sinh{(\beta (\gamma_k -\mu_d))}}{1+\cosh{(\beta(\gamma_k -\mu_d))}} = 0.\nonumber
\end{align}
Note that we use a rescaled exchange coupling of $\tilde{J} \rightarrow 0.6 \tilde{J}$ here, as proposed by Ribeiro and Wen to counteract the overemphasized AFM ordering tendency in such a mean-field analysis. The rescaling factor is motivated by experimental results on AFM ordering in cuprates.

\emph{Fractionalized Fermi liquid: $b_0=b_1 = n^z = m^z = 0$.--}
Since we focus on a SU(2) symmetric ground state as proposed for the pseudogap regime, ordering is absent and again $b_0 = b_1 =0$.  We can thereby most simply adapt the above self-consistency equations \eqref{SCE_M_1} and enforce the constraint $n^z=m^z=0$.

\bibliography{SignatureBSP}
\bibliographystyle{apsrev4-1}

\end{document}